\documentclass[aps,prd,twocolumn,groupedaddress,showpacs,showkeys]{revtex4}

\usepackage{graphicx}
\begin{document}

\newcommand{\be}{\begin{equation}}
\newcommand{\ee}{\end{equation}}

\title{Commutation relations for the electromagnetic field in the presence of dielectrics and conductors}

\author{Giuseppe Bimonte}
\email[Bimonte@na.infn.it]
\affiliation{Dipartimento di Scienze Fisiche Universit\`{a} di
Napoli Federico II Complesso Universitario MSA, Via Cintia
I-80126 Napoli Italy and INFN Sezione di Napoli, ITALY\\
}

\date{\today}

\begin{abstract}

We determine the commutation relations satisfied by the quantized
electromagnetic field  in the presence of macroscopic dielectrics
and conductors, with arbitrary dispersive and dissipative
properties. We consider in detail the case of two plane-parallel
material slabs, separated by an empty gap, and we show that at all
points in the empty region between the slabs, including their
surfaces, the electromagnetic fields always satisfy free-field
canonical equal-time commutation relations.
This result is a   consequence of general analyticity and fall-off
properties at large frequencies satisfied by the reflection
coefficients of all real materials. It is also shown that this
result  does not obtain  in the case of conductors, if the latter
are modelled as perfect mirrors. In such a case, the free-field
form of the commutation relations is recovered only at large
distances from the mirrors, in agreement with  the findings of
previous authors. Failure of perfect-mirror boundary conditions to
reproduce the correct form of the commutation relations near the
surfaces of the conductors, suggests that caution should be used
when these idealized boundary conditions are used in
investigations of proximity phenomena originating from the
quantized electromagnetic field, like the Casimir effect.

\end{abstract}


\maketitle

\section{INTRODUCTION}

The interaction  of radiation with matter has always been a
fascinating subject of investigation, and in fact it is at the
roots of quantum mechanics, with Planck's work on black body
radiation. Even though, after the development of Quantum
Electrodynamics (QED) in the middle years of last century, all
fundamental principles involved in this interaction are
undoubtedly well understood at the microscopic level, recent
experimental advances have prompted much interest in theoretical
studies of the quantized electromagnetic (e.m.) field  in close
proximity to macroscopic bodies. A thorough understanding of this
problem is indeed needed for a correct interpretation of numerous
important proximity phenomena of e.m. origin,  that include cavity
QED \cite{haroche}, the Casimir effect \cite{bordag}, radiative
heat transfer \cite{volokitin}, quantum friction \cite{kardar},
the Casimir-Polder interaction of Bose-Einstein condensates with a
substrate \cite{cornell}, etc. Apart from the intrinsic interest
of these phenomena, it has been shown recently that the quantum
fluctuations of the e.m. field surrounding macroscopic bodies,
that are at the origin of the Casimir effect, could have exciting
application in nanotechnology \cite{capasso}.

The common feature of the above e.m. phenomena, is  that they all
involve several macroscopic bodies and possibly one or more
microscopic objects (atoms, ions etc.) placed in a vacuum and
separated by distances (typical separations range from a few tens
of nanometers to several microns) that, while small from a
macroscopic point of view, are still large compared to the
interatomic distance in condensed bodies. In such circumstances,
the microscopic point of view is not of great help, because the
long range character of the e.m. field implies that
macroscopically large number of atoms are inevitably involved in
the interaction. A much more effective approach would be to
describe the influence of the macroscopic bodies on the quantized
e.m. field in the vacuum just outside their boundaries, in terms
of macroscopic features of the bodies like the electric and/or
magnetic permittivities. On physical grounds, one expects that
such an approach  should be feasible, in certain circumstances at
least, because the wavelengths of the e.m. fields participating in
these phenomena are expected to be of the order of the bodies
separations, and are therefore large on the atomic scale. This
being the case, use of   macroscopic response functions of the
bodies should be legitimate. An inevitable complication that one
faces though, when dealing with macroscopic response functions of
real bodies, is that they always display dispersion and
absorption. As is it well known, the former feature is
mathematically reflected in the fact that response functions
depend on the frequency $\omega$ (we shall neglect spatial
dispersion, and therefore we shall not consider the possible
dependence of the response functions on the wave-vector ${\bf
k}$), while the presence of dissipation entails that the response
functions have a non-vanishing imaginary part. The existence of
absorption, in particular, greatly complicates explicit
quantization of the macroscopic e.m. field. Unfortunately, such a
difficulty cannot be disposed of by simply neglecting dissipation,
because dispersive, real-valued response functions inevitably
violate causality, and must therefore be rejected.

Fortunately, though, there exists a way out that avoids the above
mentioned difficulties. This is so because a full quantization of
the e.m. field is  usually not needed,  as the   quantities of
interest are typically statistical averages of quadratic
expressions involving the macroscopic e.m. field. For systems that
are in thermodynamic equilibrium, such averages can be expressed
in terms of  (the imaginary part of) suitable macroscopic response
functions, as a result of general fluctuation-dissipation theorems
derived in the framework of linear-response theory \cite{callen}.
This general approach was probably pioneered by Rytov \cite{rytov}
in his investigations of e.m. fluctuations in the presence of
macroscopic bodies in thermal equilibrium, and it was later used
by Lifshitz \cite{lifs} in his famous theory of dispersion forces
between macroscopic condensed bodies.  In one form or another, the
fluctuation-dissipation theorem is used in all existing approaches
to problems involving the quantized e.m. field in the vicinity of
or inside macroscopic bodies. In the seventies of last century,
Agarwal used it as the basis of a systematic investigation of QED
in the presence of dielectrics and conductors \cite{agarwal}. For
a review of the most recent work we address the reader to
Refs.\cite{matloob,raabe} (see also Refs. therein). It is
important to note that this approach is not restricted to systems
in {\it global} thermodynamic equilibrium, as it is still valid in
systems that are only in {\it local} thermodynamic equilibrium.
This feature permits to include within the scope of the theory
other important phenomena, like radiative heat transfer between
closely separated bodies (for a recent review see
\cite{volokitin}), and quantum friction \cite{kardar}. Recently,
the theory has also been applied to the investigation of
Casimir-Polder \cite{cornell,buhmann} and Casimir \cite{antezza}
forces out of thermal equilibrium.

In this paper, we reexamine the basic quantum-field-theoretical
problem of the commutation relations satisfied by the quantized
e.m. field in the presence of dielectrics and/or conductors, in
the framework of the general macroscopic theory described above.
Our interest in this problem arose from a paper by Milonni
\cite{milonni} on the Casimir effect, in which it was found that
near a perfectly reflecting slab, the transverse vector potential
and the electric field satisfy a set of equal-time canonical
commutation relations of a different form form those holding for
free fields. This result is quite worrisome, in view of the very
fundamental character of commutation relations, because it
contradicts one's expectations based on microscopic theory, and
therefore it deserves detailed investigation. We remark that
unexpected commutation relations between the annihilation and
creation operators for the e.m. field inside a cavity were also
found more recently in Ref. \cite{ueda}. A partial resolution of
the paradox was offered in Ref. \cite{barnett}, in which the
problem of the e.m. commutation relations was investigated within
a simplified form of QED, in one space dimension. By relying on a
simple quantum theory of the one-dimensional lossy beam splitter,
along the lines of Ref. \cite{jaekel} (see also \cite{genet}), it
was shown that the anomalies found in \cite{ueda} in the
commutators of the annihilation and creation operators were
associated with a particular choice of the the cavity e.m. modes.
The authors of Refs. \cite{barnett}  also showed that the
canonical one-dimensional commutation relations involving the
vector potential and the electric field do not display any
anomalous behavior. No detailed explanation was however provided
for the modified form of the equal-time commutators derived in
Ref. \cite{milonni}, apart from the remark that the boundary
conditions (b.c.) satisfied by the e.m. field in the case of ideal
mirrors are incompatible with the transverse delta function form
of the full canonical commutator in three dimensions. These
authors further conjecture that standard equal-time commutators
would probably be restored after incorporating, in the full
three-dimensional setting, the physical requirements of finite
reflectivity and absorption losses by the mirrors.

Addressing this problem in detail is not only interesting as a
matter of principle, but it is also important for a better
understanding of the numerous proximity   phenomena arising from
quantum fluctuations of the e.m. field described earlier.  In many
theoretical investigations  of these phenomena, one deals with
conductors that are frequently modelled  as ideal mirrors. A
famous  example of this is provided by original Casimir's
derivation \cite{casimir} of the effect  that goes under his name.
It is then important to know to what extent conclusions drawn from
the ideal-metal model can be trusted. Indeed Casimir physics
offers examples  where predictions drawn from the ideal-metal
model are in contradiction with those derived by more realistic
modelling of the plates. One such example is still much debated as
we write, and it is the problem of determining the influence of
temperature on the magnitude of the Casimir force between two
metallic plates in vacuum. It turns out the the ideal-metal model
predicts a thermal force that, for sufficiently large separations
between the plates,  attains a magnitude   which is  twice the one
calculated on the basis of  realistic dielectric models of a
conductor, displaying a {\it finite}, though large, dc
conductivity (for a review of this puzzle, see for example
Ref.\cite{brevik} and References therein).

In order to shed light on this question, in this paper we work out
a detailed analysis of the full three-dimensional commutators for
the e.m. field, in the presence of dielectric and/or conducting
walls with arbitrary dispersion and dissipative features. The
analysis turns out to be considerably more involved than the
simple one-dimensional model studied in \cite{barnett}. Our main
result is that the canonical commutation relations satisfied by
free e.m. fields  are {\it always} valid at all points between two
macroscopic dielectric or conducting slabs, including their
surfaces, in full agreement with expectations based on the
microscopic theory for a system of charged non-relativistic
particles interacting with the e.m. field. This result is
consequence of analyticity and fall-off properties at large
frequencies of reflection coefficients of all real materials, as
it was correctly conjectured in Ref. \cite{barnett}. We also show
that such a result is not recovered, however, in the case of
conductors, if they are modelled as perfect mirrors. In this case
we find that near the conductors the equal-time commutation
relations of the  vector potential with the electric field have a
different form from the free-field case. Only at points that are
sufficiently far from the conductors, the free-fields commutators
are recovered. Our results generalize those obtained by Milonni,
in the one slab setting, and show that the modified form of the
commutation relation entailed by perfect-mirror b.c. are indeed an
artifact of these idealized b.c., not shared by real materials.

The paper is organized as follows. In Section II we recall the
basic commutation relations satisfied, within the microscopic
theory, by the e.m. field in vacuum and in the presence of charged
particles. In Section III we briefly review some general results
of linear response theory, as applied to macroscopic quantum
electrodynamics, and derive formulae for the expectation values of
the field commutators outside a system of macroscopic bodies, in
terms of suitable classical Green's functions. In Section IV we
estimate the Green's functions for a system of one or two
dielectric and/or conducting slabs in vacuum, and in Sec V we use
them to calculate the commutation relations satisfied by the e.m.
field outside the slabs. In Sec VI we consider the case of ideal,
perfectly reflecting slabs, while Section VII contains our
conclusions. Finally, three Appendices close the paper.

\section{COMMUTATION RELATIONS FOR E.M. FIELDS: MICROSCOPIC THEORY}

In this Section we briefly recall well known properties of the
commutation relations satisfied by  e.m. fields, in the framework
of a {\it microscopic} theory of non-relativistic matter, where
ponderable matter is modelled as a collection of non-relativistic
charged particles. Here and afterwards, we work in Gaussian e.m.
units, and we adopt the Coulomb gauge. As it is well known the
Coulomb gauge is very convenient for studying problems where
matter is non-relativistic, and high-energy processes are
neglected, for it allows a clear separation of electrostatic and
magnetic couplings. In this gauge, quantization is straightforward
(see for example the book \cite{cohen}). We consider first the
case of free-fields.

\subsection{Free fields}

In empty space, Maxwell equations imply that   the electric field
is purely transverse \be  {\bf
E}_{\perp}=-\frac{1}{c}\,\frac{\partial {{\bf
A}}_{\perp}}{\partial t}\;,\label{Eperp}\ee where ${\bf
A}_{\perp}$ is the transverse vector potential: \be {\bf \nabla}
\cdot {\bf A}_{\perp}={\bf 0}\;.\ee The fields ${\bf A}_{\perp}$
and ${\bf E}_{\perp}$ satisfy the following well known equal-time
canonical commutation relations: \be [{ A}_{\perp\,i}({\bf
r},t),{A}_{\perp\,j}({\bf r}',t)] = 0\;,\label{freeAA}\ee \be [{
A}_{\perp\,i}({\bf r},t),{ E}_{\perp\,j}({\bf r}',t)] = -4 \pi i
\,\hbar \,c\, \delta_{ij}^{\perp}({\bf r}-{\bf r}')\;,
\label{freeAE}\ee \be [{ E}_{\perp\,i}({\bf r},t),{
E}_{\perp\,j}({\bf r}',t)] = 0\;,\label{freeEE} \ee where
$\delta_{ij}^{\perp}({\bf x})$ is the transverse delta function
\footnote{For a review of the properties of the transverse delta
function, the reader may consult the book \cite{cohen}}: \be
\delta_{ij}^{\perp}({\bf x})=\int{d^3 {\bf k}}
\left(\delta_{ij}-\frac{k_i k_j}{k^2} \right)\,e^{i{\bf k} \cdot
{\bf x}}\,,\label{delta}\ee with $k=|{\bf k}|$.

\subsection{e.m. fields coupled to charged particles}

When charged particles are present, the phase space of the total
system includes, besides the transverse e.m. fields ${\bf
A}_{\perp}$ and ${\bf E}_{\perp}$, the positions ${\bf
x}^{(\alpha)}$, the conjugate momenta ${\bf p}^{(\alpha)}$ and the
spins ${\bf s}^{(\alpha)}$ of the particles (labelled by the index
$\alpha$). They satisfy the standard (equal-time) commutation
relations of non-relativistic Quantum Mechanics: \be
[x_i^{(\alpha)},p_j^{(\beta)}]=i \hbar \delta_{ij} \delta_{\alpha
\beta}\;,\ee \be [s_i^{(\alpha)},s_j^{(\beta)}]=i\, \hbar\,
\epsilon_{ijk}\,s_k^{(\alpha)} \delta_{\alpha \beta}\;,\ee with
all other commutators vanishing.  In particular $x_i^{(\alpha)}$,
$p_i^{(\alpha)}$ and $s_i^{(\alpha)}$ all commute with the
transverse e.m. fields ${\bf A}_{\perp}$ and ${\bf E}_{\perp}$.
Finally ${\bf A}_{\perp}$ and ${\bf E}_{\perp}$ satisfy the same
equal-time commutation relations holding in empty space,
Eqs.(\ref{freeAA}-\ref{freeEE}).

When charges are present, the electric field ${\bf E}$ has also a
longitudinal component ${\bf E}_{\|}$: \be {\bf E}={\bf
E}_{\|}+{\bf E}_{\perp}\;,\ee where ${\bf E}_{\perp}$ is still
given by Eq. (\ref{Eperp}), while ${\bf E}_{\|}$ is equal to \be
{\bf E}_{\|}({\bf r},t)=-{\bf \nabla} U({\bf r},t)\;,\ee where $U$
is the scalar potential: \be U({\bf r},t)=\sum_{\alpha}
\frac{e^{(\alpha)}}{|{\bf r}-{\bf
x}^{(\alpha)}(t)|}\;,\label{Upart}\ee with $e^{(\alpha)}$  the
charge of particle $\alpha$. The scalar potential has to be
regarded as a function of the particles positions, and therefore
it is not an independent degree of freedom of the system. Since
the particle positions commute among themselves and with the
transverse fields, ${\bf A}_{\perp}$ and ${\bf E}_{\perp}$, it
follows that \be [{ U}({\bf r},t),U({\bf
r}',t)]=0\;,\label{canUU}\ee   \be [U({\bf r},t),{
A}_{\perp\,j}({\bf r}',t)]= 0\;,\label{canUA}\ee \be [U({\bf
r},t),{ E}_{\perp\,j}({\bf r}',t)]=0\;.\label{canUE}\ee The above
Equations imply that the equal-time commutation relations
Eqs.(\ref{freeAA}-\ref{freeEE}) remain valid, irrespective of the
number and positions of the charged particles, if we replace
everywhere the transverse electric field ${ E}_{\perp\,j}$ by the
total electric field  $E_j$:\be [{ A}_{\perp\,i}({\bf
r},t),{A}_{\perp\,j}({\bf r}',t)] = 0\;,\label{freeAAbis}\ee \be
[{ A}_{\perp\,i}({\bf r},t),{E}_{j}({\bf r}',t)] = -4 \pi i
\,\hbar \,c\, \delta_{ij}^{\perp}({\bf r}-{\bf r}')\;,
\label{charAE}\ee \be [{ E_i}({\bf r},t),{ E}_{j}({\bf r}',t)] =
0\;.\label{charEE} \ee The obvious conclusion that can be drawn
from these elementary remarks is that, {\it within the microscopic
theory},  the canonical equal-time commutation relations satisfied
by the e.m. fields, Eqs. (\ref{freeAA}-\ref{freeEE}) or,
alternatively, Eqs.(\ref{freeAAbis}-\ref{charEE}) should be valid
always, and therefore they should hold, in particular, inside a
cavity made of an arbitrary material.

\section{COMMUTATION RELATIONS FOR E.M. FIELDS: MACROSCOPIC THEORY}

In this Section, we recall a few basic formulas from
linear-response theory and we discuss the type of probes that are
needed in order to obtain the commutation relations satisfied by
the {\it macroscopic} e.m. field in the presence of dielectrics
and conductors. For a review of linear-response theory we address
the reader to Refs.\cite{callen}.

In linear-response theory, one considers a quantum-mechanical
system, characterized by a (time-independent) Hamiltonian $H_0$,
in a state of thermal equilibrium described by the density matrix
$\rho$ \be \rho=e^{-\beta H}/{\rm tr}(e^{-\beta H})\;,\ee where
$\beta=1/(k_B\,T)$, with $k_B$ Boltzmann constant and $T$ the
temperature. The system is then perturbed by an external
perturbation of the form: \be H_{\rm ext}=-\int d^3{\bf r} \sum_j
Q_j({\bf r},t)\,f_j({\bf r},t)\,\ee where $f_j({\bf r},t)$ are the
external classical forces, and $Q_j({\bf r},t)$ is the dynamical
variable of the system conjugate to the force $f_j({\bf r},t)$.
One may assume, without loss of generality, that the equilibrium
values  of the quantities $Q_j({\bf r},t)$ all vanish: $\langle
Q_j({\bf r},t)\rangle=0$. The presence of the external forces
causes a deviation $\delta \langle Q_i({\bf r},t)\rangle$ of the
expectation values of $Q_j({\bf r},t)$ from their equilibrium
values. If the forces $f_j({\bf r},t)$ are sufficiently weak,
$\delta \langle Q_i({\bf r},t)\rangle$ can be taken to be linear
functionals of the applied forces $f_j({\bf r},t)$, and   one may
write: \be \delta \langle Q_i({\bf r},t)\rangle=\sum_j \int
d^3{\bf r} \int_{-\infty}^t dt' \phi_{ij}({\bf r},{\bf
r}',t-t')\,f_j({\bf r}',t')\;.\ee The above Equation assumes that
the system was in equilibrium at $t=-\infty$, and that it reacts
to the  external force in a causal way. The quantities
$\phi_{ij}({\bf r},{\bf r}',t-t')$ are called response functions
of the system. In principle, they can be measured by applying to
the system of interest suitable external classical probes.

By a straightforward computation in time-dependent perturbation
theory one may prove that the response functions $\phi_{ij}({\bf
r},{\bf r}',t-t')$  are related to the equilibrium (i.e. in the
absence of the external forces) expectation values  of the
commutators of the dynamical variables $Q_i({\bf r},t)$: \be
\phi_{ij}({\bf r},{\bf r}',t-t')=\frac{i}{ \hbar}\,\langle
[Q_i({\bf r},t),Q_j({\bf
r}',t')]\;\rangle\;\theta(t-t'),\label{resp}\ee where $\theta(x)$
is Heaviside step function ($\theta(x)=1$ for $x > 0$,
$\theta(x)=0$ for $x<0$) and $Q_i({\bf r},t)$ is the Heisenberg
operator: \be Q_i({\bf r},t)=e^{i H_0 t/\hbar} Q_i({\bf r},0)e^{-i
H_0 t/\hbar}\;.\ee As it is well known, Eq. (\ref{resp}) is the
starting point from which several general fluctuation-dissipation
theorems can be derived, that allow to express the (symmetrized)
correlation functions of the quantities $Q_i({\bf r},t)$ in terms
of the dissipative component of the response functions
$\phi_{ij}$. Since we shall not make use of these theorems in what
follows, we shall not present them here, and we address the
interested reader to Refs.\cite{callen} for details.

We wish to exploit Eq. (\ref{resp}) to study the commutation
relations satisfied by the {\it macroscopic} e.m. field at points
placed outside a number of dielectric or conducting bodies. For
this purpose, following Agarwal \cite{agarwal}, we take the
external probes to be a system of classical   electric and
magnetic dipoles, with densities ${\bf P}({\bf r},t)$ and ${\bf
M}({\bf r},t)$ respectively, placed outside the bodies. The
external Hamiltonian $H_{\rm ext}$ is then   of the form: \be
H_{\rm ext}=-\int d^3{\bf r} [{\bf P}^{({\rm ext})}({\bf r},t)
\cdot {\bf E}({\bf r},t)+{\bf M}^{({\rm ext})}({\bf r},t) \cdot
{\bf B}({\bf r},t)]\;.\label{hext0}\ee It is convenient for our
purposes to have distinct probes for the longitudinal and the
transverse components of the e.m. field. This can be achieved by
demanding that ${\bf P}^{({\rm ext})}$ be curl free
\be {\bf \nabla} \times {\bf P}^{({\rm ext})}={\bf 0}\;,
\;.\label{sour}\ee If we now express in Eq. (\ref{hext0}) the e.m.
field in terms of the scalar and vector potentials: \be {\bf
E}=-{\bf \nabla} U-\frac{1}{c}\frac{\partial {{\bf
A}}_{\perp}}{\partial t}\;,\;\;\;{\bf B}={\bf \nabla}\times {\bf
A}_{\perp}\;,\ee after an integration by parts, and exploiting Eq.
(\ref{sour}), the external Hamiltonian can be rewritten as: \be
H_{\rm ext}=\int d^3{\bf r}[U({\bf r},t) \rho^{({\rm ext})} ({\bf
r},t)-\frac{1}{c} {\bf A}_{\perp}({\bf r},t)\cdot {\bf j}^{({\rm
ext})}_{\perp} ({\bf r},t) ]\;, \label{Hextfin}\ee where
$\rho^{({\rm ext})} =-{\bf \nabla} \cdot {\bf P}^{({\rm ext})}$
and ${\bf j}^{({\rm ext})}_{\perp}=c \,{\bf \nabla} \times {\bf
M}^{({\rm ext})}$. Note that the current ${\bf j}^{({\rm
ext})}_{\perp}$ is transverse: \be {\bf \nabla} \cdot {\bf
j}^{({\rm ext})}_{\perp} =0\;.\ee We remark once again that the
scalar potential $U({\bf r},t)$, in the external Hamiltonian Eq.
(\ref{Hextfin}) does not represent an independent dynamical
variable, and it must be regarded as a function of the particle's
position, according to Eq.(\ref{Upart}). Therefore, in the absence
of matter, no such term is present in the external Hamiltonian,
and the scalar potential is zero.

The response functions are then computed by solving the classical
{\it macroscopic} Maxwell Equations with $\rho^{({\rm ext})}$ and
${\bf j}^{({\rm ext})}_{\perp}$ as external sources: \be {\bf
\nabla} \cdot {\bf D}=4 \pi \rho^{({\rm ext})}\;,\label{max1}\ee
\be {\bf \nabla} \times {\bf E}+\frac{1}{c}\frac{\partial {\bf
B}}{\partial t}=0\;,\label{max2}\ee \be {\bf \nabla} \cdot {\bf
B}=0\;, \label{max3}\ee \be {\bf \nabla} \times {\bf
H}=\frac{1}{c}\frac{\partial {\bf D}}{\partial t}+\frac{4 \pi}{c}
\,{\bf j}^{({\rm ext})}_{\rm tot}\;,\label{max4}\ee where \be {\bf
j}^{({\rm ext})}_{\rm tot}={\bf j}^{({\rm
ext})}_{\perp}+\frac{\partial {\bf P}^{({\rm ext})}}{\partial
t}\;.\ee The above equations have to be solved subject to the
usual  b.c. of macroscopic electrodynamics, namely (i) tangential
components of ${\bf E}$ and ${\bf H}$ and (ii) normal components
of ${\bf D}$ and ${\bf B}$ must be continuous across the bodies
interfaces, which are assumed to have sharp boundaries. Coherently
with the spirit of a macroscopic approach, the dielectrics and the
conductors will be described in terms of the appropriate electric
and magnetic susceptibilities. We suppose from now on that the
bodies are made of non-magnetic ($\mu=1$)), isotropic and
spatially non-dispersive materials, characterized by a
frequency-dependent electric permittivity $\epsilon({\bf
r},\omega)$. We also assume that the bodies are homogeneous, in
such a way that the permittivity $\epsilon({\bf r},\omega)$ is
independent of ${\bf r}$ within the volume occupied by each body,
with discontinuities occurring only at the bodies interfaces.
By virtue of   homogeneity of the bodies, and of linearity of the
b.c. at the bodies interface, the field Equations for the scalar
potential $U({\bf r},t)$ are completely decoupled from those for
the transverse vector potential ${\bf A}({\bf r},t)$. Therefore,
we have two independent sets of Green's functions: \be {U}({\bf
r},t)=\int_{-\infty}^{t} d t' \int d^3 {\bf r}' {{G}}({\bf r},{\bf
r}',t-t')\,{\rho}^{({\rm ext})}({\bf r}',t')\;,\ee \be {{\bf
A}}_{\perp}({\bf r},t)=\frac{1}{c}\int_{-\infty}^{t}  d t'\int d^3
{\bf r}' { {\bf G}}_{\perp}({\bf r},{\bf r}',t-t')\cdot{ {\bf
j}}_{\perp}^{({\rm ext})}({\bf r}',t')\;,\ee where ${ {\bf
G}}_{\perp}({\bf r},{\bf r}',t-t')$ has to be understood as a
dyadic Green function.

From the general result of linear-response theory, Eq.
(\ref{resp}), we then obtain the following expressions for the
two-times expectation values of the commutators of the e.m.
potentials: \be \langle[U({\bf r},t),U({\bf r}',t')]\rangle=i
\,\hbar\, G({\bf r},{\bf r}',t-t') \,,\label{gencomUU}\ee \be
\langle[U({\bf r},t),A_{\perp i}({\bf
r}',t')]\rangle=0\,,\label{gencomUA}\ee \be \langle[A_{\perp
i}({\bf r},t),A_{\perp j}({\bf r}',t')]\rangle=-i \,\hbar\,
G_{\perp ij}({\bf r},{\bf r}',t-t')\,,\label{gencomAA}\ee  where
$t > t'$. For our purposes, it is convenient to split the Green's
functions, {\it outside} the bodies, as sums of an empty-space
contribution plus a correction arising from the material bodies:
\be {  { G}} ({\bf r},{\bf r}',t-t')= {{ G}}^{(0)}({\bf r}-{\bf
r}',t-t')+\,{{F}}^{(\rm mat)}({\bf r},{\bf
r}',t-t')\;,\label{decscagen}\ee and \be {{\bf G}}_{\perp}({\bf
r},{\bf r}',t-t')={{\bf G}}^{(0)}_{\perp}({\bf r}-{\bf r}',t-t')+{
{\bf F}}^{(\rm mat)}_{\perp}({\bf r},{\bf
r}',t-t')\;.\label{dectengen}\ee Here, ${{ G}}^{(0)}$ and ${{\bf
G}}^{(0)}_{\perp}$ denote the Green's functions in free space,
while ${{F}}^{(\rm mat)}$ and ${{\bf F}}^{(\rm mat)}_{\perp}$
describe the effects resulting from the presence of the bodies.
Such a splitting presents the advantage that all singularities are
included in the free parts ${{ G}}^{(0)}$ and ${{\bf
G}}^{(0)}_{\perp}$, while the quantities ${{F}}^{(\rm mat)}$ and
${{\bf F}}^{(\rm mat)}_{\perp}$ are smooth ordinary functions of
${\bf r}$ and ${\bf r}'$. The free-field Green's functions have
the following well-known expressions: \be {{ G}}^{(0)}({\bf
r}-{\bf r}') =\frac{1}{|{\bf r}-{\bf r}'|}\,\delta(t-t')\;,\ee and
\be { {G}}^{(0)}_{\perp ij}=c\,\int \frac{d^3 {\bf k}}{2 \pi^2\,k}
\left( \delta_{ij}-\frac{k_i\,k_j}{k^2}\right) \;e^{i{\bf k}\cdot
({\bf r}-{\bf r}')}\sin[kc(t-t')]\;.\ee The factor $\delta(t-t')$
in the expression of ${ G}^{(0)}$ expresses the instantaneous
character of the  longitudinal electric field in the Coulomb
gauge. The expressions for the equal-time commutators of the e.m.
fields are easily derived by taking suitable limits of Eqs.
(\ref{gencomUU}-\ref{gencomAA}) and of their time derivatives, for
$t \rightarrow t'^+$.     Upon using Eqs. (\ref{decscagen}) and
(\ref{dectengen}), and exploiting the following three relations
that are obvious consequences of Eq. (\ref{Eperp}):
\be\langle[U({\bf r},t),E_{\perp i}({\bf
r}',t')]\rangle=-\frac{1}{c}\frac{\partial}{\partial
t'}\langle[U({\bf r},t),A_{\perp i}({\bf
r}',t')]\rangle\,,\label{comcavAEperp}\ee \be\langle[A_{\perp
i}({\bf r},t),E_{\perp j}({\bf
r}',t')]\rangle=-\frac{1}{c}\frac{\partial}{\partial
t'}\langle[A_{\perp i}({\bf r},t),A_{\perp j}({\bf
r}',t')]\rangle\,,\label{comcavAEperp}\ee and \be\langle[E_{\perp
i}({\bf r},t),E_{\perp j}({\bf
r}',t')]\rangle=\frac{1}{c^2}\frac{\partial^2}{\partial
t\,\partial t'}\langle[A_{\perp i}({\bf r},t),A_{\perp j}({\bf
r}',t')]\rangle\,,\label{comcavEEperp}\ee from Eqs.
(\ref{gencomUU}-\ref{gencomAA}) we obtain: \be  \langle[U({\bf
r},t),U({\bf r}',t)]\rangle = i \,\hbar\, \;A^{(\rm mat)}({\bf
r},{\bf r}')\,,\label{etcomcavUU}\ee \be  \langle[U({\bf
r},t),A_{\perp i}({\bf r}',t)]\rangle =0\,,\label{etcomcavUA} \ee
\be  \langle[U({\bf r},t),E_{\perp i}({\bf r}',t)]\rangle =
0\,,\label{etcomcavUE}\ee \be
  \langle[A_{\perp
i}({\bf r},t),A_{\perp j}({\bf r}',t)]\rangle = -i
\,\hbar\,A_{\perp ij}^{({\rm mat})}({\bf r},{\bf r}')
\,,\label{etcomcavAA} \ee $$\langle[A_{\perp i}({\bf
r},t),E_{\perp j}({\bf r}',t)]\rangle$$ \be=-4 \pi i \,\hbar \,c\,
\delta_{ij}^{\perp}({\bf r}-{\bf r}')+\;
\,\frac{i\,\hbar}{c}\,B_{\perp ij}^{({\rm mat})}({\bf r},{\bf r}')
\,,\label{etcomcavAE}\ee and \be \langle[E_{\perp i}({\bf
r},t),E_{\perp j}({\bf r}',t)]\rangle=
-\,\frac{i\,\hbar}{c^2}\,C_{\perp ij}^{({\rm mat})}({\bf r},{\bf
r}') \,,\label{etcomcavEE}\ee where we defined \be A^{(\rm
mat)}({\bf r},{\bf r}') \equiv \lim_{t \rightarrow t'^+} F^{(\rm
mat)}({\bf r},{\bf r}',t-t')\,,\label{Amat}\ee \be A_{\perp
ij}^{({\rm mat})}({\bf r},{\bf r}')\equiv  \lim_{t \rightarrow
t'^+}F_{\perp ij}^{({\rm mat})}({\bf r},{\bf
r}',t-t')\;,\label{Aijmat}\ee \be B_{\perp ij}^{({\rm mat})}({\bf
r},{\bf r}')\equiv \lim_{t \rightarrow t'^+}\frac{
\partial F_{\perp ij}^{({\rm mat})}}{\partial t'}({\bf r},{\bf
r}',t-t')\;,\label{Bijmat}\ee and \be C_{\perp ij}^{({\rm
mat})}({\bf r},{\bf r}')\equiv  \lim_{t \rightarrow t'^+}\frac{
\partial^2 F_{\perp ij}^{({\rm mat})}}{\partial t \,\partial t'}({\bf r},{\bf
r}',t-t') \,. \label{Cijmat}\ee  By comparing Eqs.
(\ref{etcomcavUU}-\ref{etcomcavEE}) with
Eqs.(\ref{freeAA}-\ref{freeEE}) and Eqs.
(\ref{canUU}-\ref{canUE}), we see that outside the bodies the
free-field canonical commutation relations are recovered provided
that the quantities $A^{(\rm mat)}$, $A_{\perp ij}^{({\rm mat})}$,
$B_{\perp ij}^{({\rm mat})}$ and $C_{\perp ij}^{({\rm mat})}$  are
zero. We shall prove below that this indeed the case, as a result
of analyticity and fall off properties  at large frequencies of
the reflection coefficients of all real materials.

Before  we turn to detailed computations, we present below the
field Equations satisfied by ${G}$ and ${\bf G}_{\perp}$. They are
conveniently expressed in terms of the (one-sided) Fourier
transforms of the Green's functions, defined as: \be
\tilde{G}({\bf r},{\bf r}',\omega)=\int_0^{\infty} dt \,{G}({\bf
r},{\bf r}',t)\,e^{i \omega t}\;, \ee \be \tilde{{\bf
G}}_{\perp}({\bf r},{\bf r}',\omega)=\int_0^{\infty} dt \,{\bf
G}_{\perp}({\bf r},{\bf r}',t)\,e^{i \omega t}\;.\ee From Maxwell
Equations we then obtain: \be {\bf \nabla}\cdot [\,\epsilon({\bf
r},\omega) \,{\bf \nabla} \tilde{G}\,]=-4 \pi \,\delta({\bf
r}-{\bf r'})\;,\label{diffsca}\ee \be (\triangle + \,\epsilon({\bf
r},\omega)\,\omega^2/c^2){\tilde {\bf G}}_{\perp}({\bf r},{\bf
r}',\omega)=- {4 \pi} \,{\bf \delta}_{\perp}({\bf r}-{\bf
r'})\;,\label{difften}\ee where ${\bf \delta}_{\perp}({\bf r}-{\bf
r'})$ is the transverse delta-function dyad, Eq. (\ref{delta}).
These Equations must be solved with the appropriate b.c. at the
bodies interfaces, and must be subject to the conditions required
for a retarded Green's function \cite{abrikosov}. For later use,
it is useful to recall the main properties enjoyed by the Green's
functions \cite{eckhart}. First of all, they satisfy the following
{\it reciprocity} relations: \be {\tilde {G}}({\bf r},{\bf
r}',\omega)={\tilde {G}}({\bf r}',{\bf r},\omega)\;,\ee and
\be{\tilde {G}}_{\perp ij}({\bf r},{\bf r}',\omega)={\tilde
{G}}_{\perp ji}({\bf r}',{\bf r},\omega)\;,\ee that are a
consequence of microscopic reversibility. The next set of
properties express  reality features of the Green' functions, and
are a direct consequence of reality of the external sources
\be{\tilde {G}}^*({\bf r},{\bf r}',\omega)={\tilde {G}}({\bf
r},{\bf r}',-\omega)\;,\label{rescauno}\ee and \be{\tilde
{G}}^*_{\perp ij}({\bf r},{\bf r}',\omega)={\tilde {G}}_{\perp
ij}({\bf r},{\bf r}',-\omega)\;.\label{retenuno}\ee The next set
of properties is a consequence of  the fact that the  permittivity
$\epsilon(\omega)$ of any causal medium is an analytic function of
the frequency $w$ in the upper complex half-plane ${\cal C}^+$
\cite{lifs2} (see also Appendix B)). This implies that the Green's
functions ${\tilde {G}}({\bf r},{\bf r}',\omega)$ and ${\tilde
{\bf G}}_{\perp}({\bf r},{\bf r}',\omega)$  are also {\it
analytic} in ${\cal C}^+$, as it must be case for a retarded
response function. In ${\cal C}^+$ they satisfy the conditions
${\tilde {G}}^*({\bf r},{\bf r}',w)={\tilde {G}}({\bf r},{\bf
r}',-w^*)$ and ${\tilde {G}}^*_{\perp ij}({\bf r},{\bf
r}',w)={\tilde {G}}_{\perp ij}({\bf r},{\bf r}',-w^*)$, that
generalize the reality conditions Eq. (\ref{rescauno}) and Eq.
(\ref{retenuno}), respectively. These more general properties
imply that the Green's functions are real along the imaginary
frequency axis: \be{\tilde {G}}({\bf r},{\bf r}',i \xi)={\tilde
{G}}^*({\bf r},{\bf r}',i \xi)\;,\label{resca}\ee and \be{\tilde
{G}}_{\perp ij}({\bf r},{\bf r}',i \xi)={\tilde {G}}^*_{\perp
ij}({\bf r},{\bf r}',i \xi)\;.\label{reten}\ee It is finally
useful to write down the inversion formulas expressing the Green's
functions, in the time domain, in terms of their Fourier
transforms. They are \be { G} ({\bf r},{\bf
r}',t-t')=\int_{\Gamma} d w \, {\tilde { G}} ({\bf r},{\bf
r}',w)\,e^{-i w (t-t')}\ee \be {{\bf G}}_{\perp}({\bf r},{\bf
r}',t-t')=\int_{\Gamma} d w \, {\tilde {\bf G}}_{\perp}({\bf
r},{\bf r}',w)\,e^{-i w (t-t')}\;\label{invtra}, \ee where
$\Gamma$ is any contour in ${\cal C}^+$ that can be obtained by
smoothly deforming the real frequency axis, keeping fixed the
end-points at infinity. Analyticity of the Green's functions in
${\cal C}^+$ ensures that the integrals on the r.h.s. are
independent of the chosen contour $\Gamma$.


In the next two sections we shall compute the Green functions at
points outside  a single dielectric slab, and between two plane
parallel slabs.

\section{GREEN'S FUNCTIONS OUTSIDE DIELECTRICS AND CONDUCTORS}

In this Section we evaluate the e.m. Green's functions outside
dielectric and/or conducting slabs.

In the next two subsections we shall separately consider the cases
of one slab in vacuum, and two plane-parallel slabs separated by
an empty gap. We choose our cartesian coordinate system such that
the $z$-axis is perpendicular to the slabs. Translational
invariance of the system in the $(x,y)$ plane implies that the
quantities ${\tilde {F}}^{(\rm mat)}$ and ${\tilde {\bf F}}^{(\rm
mat)}_{\perp}$ are functions only of $z$, $z'$ and $({\bf
r}_{\perp}-{\bf r}_{\perp}')$, where we denote by ${{\bf
x}_{\perp}}$ the projection of the vector ${\bf x}$ onto the
$(x,y)$ plane. The computation  is facilitated if we express
${\tilde { G}}^{(0)}$ and ${\tilde {\bf G}}^{(0)}_{\perp}$   in a
form that is adapted to the symmetries of our problem. Consider
first  the free scalar Green's function ${\tilde { G}}^{(0)}$: \be
{\tilde { G}}^{(0)}({\bf r}-{\bf r}') =\frac{1}{|{\bf r}-{\bf
r}'|}\;. \ee We note that ${\tilde { G}}^{(0)}$ is independent of
the complex frequency  $w$, as it must be because of the
instantaneous character of the  longitudinal electric field in the
Coulomb gauge. For our purposes, the convenient form of ${\tilde {
G}}^{(0)}$ is the following well-known Weyl representation: \be
{\tilde { G}}^{(0)}= \int \frac{d^2 {\bf k}_{\perp}}{2
\pi\,k_{\perp}} e^{i {\bf k}_{\perp}\cdot ({\bf r}_{\perp}-{\bf
r}_{\perp}')-k_{\perp}|z-z'|}\;,\label{weyl}\ee that can be easily
obtained by integrating over $k_3$ the standard plane-wave
decomposition of ${\tilde { G}}^{(0)}$. The above expression for
${\tilde { G}}^{(0)}$ can also be written as  \be {\tilde {
G}}^{(0)}= \int \frac{d^2 {\bf k}_{\perp}}{2 \pi\,k_{\perp}} e^{i
{\bar {\bf k}}^{(\pm)}\cdot ({\bf r}-{\bf
r}')}\;,\label{Gzero4}\ee where we define ${\bar {\bf
k}}^{(\pm)}={\bf k}_{\perp} \pm i k_{\perp}\,{\hat {\bf z}}$, and
the upper (lower) sign is for $z \ge z'$ ($z \le z'$). Consider
now the familiar representation of ${\tilde {\bf
G}}^{(0)}_{\perp}$: \be {\tilde {G}}^{(0)}_{\perp ij}=\int
\frac{d^3 {\bf k}}{2 \pi^2} \frac{1}{k^2-k_0^2}\left(
\delta_{ij}-\frac{k_i\,k_j}{k^2}\right) \;e^{i{\bf k}\cdot ({\bf
r}-{\bf r}')}\;, \label{gthree1}\ee where $k_0=w/c$. In Appendix
A, we  show that ${\tilde {\bf G}_{\perp}}^{(0)}$ can be
decomposed as the sum of two dyads: \be {\tilde {\bf
G}_{\perp}}^{(0)}={\tilde {\bf U}}^{(0)}+{\tilde {\bf
V}}^{(0)}\;.\label{Gzerosplit}\ee Here, ${\tilde {\bf U}}^{(0)}$
denotes the tensor of components \be {\tilde { U}_{ ij}}^{(0)}=i
\int \frac{d^2 {\bf k}_{\perp}}{2 \pi\,{k_z}} \left( e_{\perp i}
e_{\perp j}+\frac{\xi^{(\pm)}_i \xi^{(\pm)}_j}{k_0^2} \right)
\,e^{i {\bf k}^{(\pm)}\cdot ({\bf r}-{\bf r}')
}\;,\label{Ufinte}\ee where $k_z=\sqrt{k_0^2-k_{\perp}^2}$ (the
square root is defined such that ${\rm Im}(k_z) > 0$), ${\bf
e_{\perp}}={\hat {\bf z}}\times {\hat {\bf k}}_{\perp}$, ${\bf
k}^{(\pm)}={\bf k}_{\perp}\pm k_z {\hat {\bf z}}$ and ${\bf
\xi}^{\pm}=k_{\perp}{\hat {\bf z}} \mp k_z {\hat {\bf
k}}_{\perp}$. As to ${\tilde {V}_{ ij}}^{(0)}$, it can be written
as \be {\tilde {V}_{ ij}}^{(0)}=\frac{1}{k_0^2} \frac{\partial^2
{\tilde \Psi}^{(0)}}{\partial x_i
\partial x'_j}\;,\label{Vfin}\ee where ${\tilde\Psi}^{(0)}$ is the
function \be {\tilde \Psi}^{(0)}=\int \frac{d^2 {\bf
k}_{\perp}}{{2 \pi}\,k_{\perp}}
 \,
 \,e^{i {\bar {\bf k}}^{(\pm)}\cdot ({\bf r}-{\bf
r}')}\;.\label{Psifin}\ee In both Eqs. (\ref{Ufinte}) and
(\ref{Psifin}) the upper (lower) sign is for $z \ge z'$ ($z \le
z'$). It is useful to provide a simple intuitive interpretation
for the above Green's functions  that will be useful later when we
consider the influence of a material slab. Consider first the
expression for ${\tilde G}^{(0)}$ given in Eq. (\ref{Gzero4}): we
can interpret is as consisting of a superposition of instantaneous
scalar waves originating from point ${\bf r}'$, that propagate to
the right (left) with wave-vector ${\bar {\bf k}}^{(+)}$ (${\bar
{\bf k}}^{(-)}$). Consider now our expression for ${\tilde {\bf
G}}^{(0)}_{\perp}$, Eq.(\ref{Gzerosplit}). Its first contribution
${\tilde {\bf U}}^{(0)}$, Eq. (\ref{Ufinte}), can be physically
interpreted as a superposition of e.m. waves with TE and TM
polarization corresponding, respectively, to the first and second
term between the round brackets in Eq. (\ref{Ufinte}). These waves
originate from point ${\bf r}'$ and propagate to the right (left)
with wave-vector ${\bf k}^{(+)}$ (${\bf k}^{(-)}$). We note that
for $k_0 > k_{\perp}$ these modes represent propagating waves,
while for $k_0 < k_{\perp}$ they are evanescent waves, that decay
exponentially as we move away from $z'$.  The second contribution
to ${\tilde {\bf G}}^{(0)}_{\perp}$, ${\tilde {\bf V}}^{(0)}$, can
instead be interpreted as representing scalar waves that propagate
instantaneously from point ${\bf r}'$ in the right (left)
direction, with wave-vector ${\bar {\bf k}}^{(+)}$ (${\bar {\bf
k}}^{(-)}$).

We are now ready to compute ${\tilde {F}}^{(\rm bodies)}$ and
${\tilde {\bf F}}^{(\rm bodies)}_{\perp}$. We consider first the
one-slab case.

\subsection{The case of one slab}

In this Section we compute the Green's functions outside a single
dielectric or conducting slab, occupying the half-space $z<0$.
Following the remarks of the previous Section, outside the slab
and on its surface, i.e. for $z,z' \ge 0$ we define: \be {\tilde {
G}}^{(\rm wall)}({\bf r},{\bf r}',w)= {\tilde { G}}^{(0)}({\bf
r}-{\bf r}')+\,{\tilde {F}}^{(\rm wall)}({\bf r}_{\perp}-{\bf
r}_{\perp}',z,z',w)\;,\label{decsca}\ee and \be {\tilde {\bf
G}}^{(\rm wall)}_{\perp}({\bf r},{\bf r}',w)={\tilde {\bf
G}}^{(0)}_{\perp}({\bf r}-{\bf r}',w)+{\tilde {\bf F}}^{(\rm
wall)}_{\perp}({\bf r}_{\perp}-{\bf
r}_{\perp}',z,z',w)\;.\label{decten}\ee

Fixing once and for all $z' \ge 0$,  we make for ${\tilde
{F}}^{(\rm wall)}$ the following ansatz: \be {\tilde {F}}^{(\rm
wall)}=- \int \frac{d^2 {\bf k}_{\perp}}{2 \pi\,k_{\perp}}\, {\bar
r}(w)\, e^{i {\bf k}_{\perp}\cdot ({\bf r}_{\perp}-{\bf
r}_{\perp}')-k_{\perp}(z+z')}\;,\;\;z \ge 0\,.\label{Fwall0}\ee
For $z<0$, the complete Green's function is taken to be of the
form: \be {\tilde { G}}^{(\rm wall)} = \int \frac{d^2 {\bf
k}_{\perp}}{2 \pi\,k_{\perp}}\, {\bar t}(w) \, e^{i {\bf
k}_{\perp}\cdot ({\bf r}_{\perp}-{\bf
r}_{\perp}')-k_{\perp}(z'-z)}\;,\;z<0\;.\ee Both ansatz ensure
appropriate fall off for $|z| \rightarrow \infty$. It is easy to
verify that the above ansatz satisfy the b.c. at $z=0$, provided
that we take: \be {\bar
r}(w)=\frac{\epsilon(w)-1}{\epsilon(w)+1}\,\label{refsca}\ee and
\be {\bar t}(w)=1-{\bar r}(w)\;. \ee The chosen forms of ${\tilde
{F}}^{(\rm wall)}$, for $z>0$, and ${\tilde { G}}^{(\rm wall)}$,
for $z<0$, have a simple physical interpretation, that will be be
useful later when we shall consider the more elaborate case of two
slabs. In empty space, the source ${\tilde \rho}({\bf r}',w)$
generates "instantaneous" scalar waves of (complex) frequency
${w}$ originating at ${\bf r}'$ and propagating in the right
direction (i.e. towards larger $z$) with (complex) wave-vector
${\bar {\bf k}}^{(+)}$, and in the left direction with wave-vector
${\bar {\bf k}}^{(-)}$. When a wall is present, the left-moving
waves hit the wall and then we have a reflected wave with
amplitude ${\bar r}(w)$, and a transmitted wave of amplitude
${\bar t}(w)$.

We can now evaluate ${\tilde {\bf F}}^{(\rm wall)}_{\perp}$. In a
way analogous to Eq. (\ref{Gzerosplit}), we  decompose it as: \be
{\tilde {\bf F}_{\perp}}^{({\rm wall})}={\tilde {\bf U}}^{({\rm
wall})}+{\tilde {\bf V}}^{({\rm wall})}\;.\label{Gwallsplit}\ee
Inside the slab, for the full Green's function we set instead: \be
{\tilde {\bf G}^{(\rm wall)}_{\perp}}={\tilde {\bf U}}^{({\rm
in})}+{\tilde {\bf V}}^{({\rm
in})}\;,\;\;\;z<0.\label{Ginsplit}\ee Linearity of the
boundary-value problem permits to determine separately ${\tilde
{\bf U}}^{({\rm wall})}$ and ${\tilde {\bf V}}^{({\rm wall})}$.
The physical picture of ${\tilde {\bf U}}^{(0)}$ as a
superposition of TE and TM waves, suggests at once the following
ansatz for ${\tilde {\bf U}}^{({\rm wall})}$:
$$ {\tilde { U}_{ij}}^{({\rm wall})}=i \int \frac{d^2 {\bf
k}_{\perp}}{2 \pi\,{k_z}} \left( e_{\perp i} e_{\perp
j}\,r^{(s)}(w,{ k}_{\perp})\right.$$ \be
 +\left.\frac{\xi^{(+)}_i \xi^{(-)}_j}{k_0^2}\,r^{(p)}(w,{
k}_{\perp}) \right) \,e^{i ({\bf k}^{(+)}\cdot {\bf r}-{\bf
k}^{(-)}\cdot {\bf r}') }\;,\label{Ufin}\ee where $r^{(s)}(w,{\bf
k}_{\perp})$ and $r^{(p)}(w,{\bf k}_{\perp})$ are the familiar
Fresnel reflections coefficients for TE and TM waves,
respectively: \be r^{(s)}(w,{
k}_{\perp})=\frac{k_z-q}{k_z+q}\;,\label{rs}\ee \be r^{(p)}(w,{
k}_{\perp})=\frac{\epsilon(w) k_z-q}{\epsilon(w)
k_z+q}\;,\label{rp}\ee where $q=\sqrt{\epsilon(w)
k_0^2-k_{\perp}^2}$. A somewhat lengthy solution of the
boundary-value problem indeed confirms the above intuitive form of
${\tilde { U}_{ij}}^{({\rm wall})}$. Consider now  ${\tilde {\bf
V}}^{({\rm wall})}$. Eq. (\ref{Psifin}) suggests that we set: \be
{\tilde {V}}_{ij}^{({\rm wall})}=\frac{1}{k_0^2} \frac{\partial^2
{ \Psi}^{({\rm wall})}}{\partial x_i \partial
x'_j}\;,\label{Vwall}\ee while inside the slab (i.e. for $z<0$) we
set: \be {\tilde {V}}_{ij}^{(\rm in)}=\frac{1}{k_0^2}
\frac{\partial^2 { \Psi}^{({\rm in})}}{\partial x_i \partial
x'_j}\;.\label{Vin}\ee It can be seen that the appropriate
boundary dielectric conditions at $z=0$ are satisfied, provided
that the   functions ${\tilde \Psi}^{({\rm 0})}$, ${\tilde
\Psi}^{({\rm wall})}$ and ${\tilde \Psi}^{({\rm in})}$ fulfill
there the following b.c.: \be {\tilde \Psi}^{({\rm
in})}|_{z=0}=({\tilde \Psi}^{({ 0})}+{\tilde \Psi}^{({\rm
wall})})|_{z=0}\;,\ee \be \epsilon(w) {\tilde \Psi}^{({\rm
in})'}|_{z=0}=({\tilde \Psi}^{({ 0})'}+{\tilde \Psi}^{({\rm
wall})'})|_{z=0}\;,\ee where a prime denotes a derivative with
respect to $z$. One then finds: \be {\tilde {\Psi}}^{(\rm wall)}=-
\int \frac{d^2 {\bf k}_{\perp}}{2 \pi\,k_{\perp}}\, {\bar r}(w)\,
e^{i {\bar {\bf k}}^{(+)}\cdot {\bf r} -i {\bar {\bf
k}}^{(-)}\cdot {\bf r}'}\;,\label{Psiwall0}\ee where ${\bar
r}(\omega)$ is the reflection coefficient in Eq. (\ref{refsca}).
We note that the expression of ${\tilde {\Psi}}^{(\rm wall)}$
coincides with that of ${\tilde {F}}^{(\rm wall)}$. We remark that
${\tilde {F}}^{(\rm wall)}$ and ${\tilde {\bf F}}^{(\rm
wall)}_{\perp}$ are analytic functions of the frequency $w$ in the
upper complex plane ${\cal C}^+$, as a result of analyticity in
${\cal C}^+$ of the reflection coefficients ${\bar r}(w)$,
$r^{(\alpha)}(w)$ (see Appendix B). Moreover, we note that
${\tilde { F}}^{(\rm wall)}$  has no singularities along the
real-frequency axis, as it can be easily checked from Eq.
(\ref{Fwall0}), if one considers that the reflection coefficient
${\bar r}(w)$ is finite in ${\cal C}^+$ (see Appendix B). As to
${\tilde {\bf F}}^{(\rm wall)}_{\perp}$, it only has an integrable
singularity at $k_z=0$. The presence of singular factors
proportional to $k_0^{-2}$ in the expressions of ${\tilde
{U}}_{ij}^{(\rm wall)}$  and ${\tilde {V}}_{ij}^{(\rm wall)}$ (see
Eq. (\ref{Ufin}) and   Eq. (\ref{Vwall})) does not cause any
further singularities at $w=0$, for it can be verified that these
singular terms cancel each other upon taking the sum of ${\tilde
{U}}_{ij}^{(\rm wall)}$ and ${\tilde {V}}_{ij}^{(\rm wall)}$, as
we now show. Indeed, upon collecting in Eq. (\ref{Ufin}) and Eq.
(\ref{Vwall}) the terms that are singular at $w=0$, we obtain:
$$\lim_{\omega\rightarrow 0}{\tilde F}^{(\rm wall)}_{\perp ij}({\bf
r},{\bf r}')) $$ \be= c^2 \lim_{w\rightarrow 0}\int \frac{d^2 {\bf
k}_{\perp}}{{2 \pi}{k_{\perp}}}\, \frac{  {r^{(p)}} - {{\bar r}}
}{w^2}\,
 {\bar k}^{(+)}_i {\bar
k}^{(-)}_j e^{i {\bar {\bf k}}^{(+)}\cdot {\bf r}-i {\bar {\bf
k}}^{(-)}\cdot{\bf r}'}  \;,\label{respm}\ee where we made use of
the following relations \be {\bf \xi}^{(\pm)}=\mp i\,{\bar {\bf
k}}^{(\pm)}+ O(w^2)\;,\ee \be {\bf k}^{(\pm)}= {\bar {\bf
k}}^{(\pm)}+ O(w^2)\;,\ee \be k_z=i\,k_{\perp}+ O(w^2)\;,\ee to
substitute everywhere ${\bf \xi}^{(\pm)}$, ${\bf k}^{(\pm)}$ and
$k_z$ by $\mp i\,{\bar {\bf k}}^{(\pm)}$, ${\bar {\bf k}}^{(\pm)}$
and $k_{\perp}$ respectively. Now  in Appendix B it is shown that
both, for dielectrics and conductors, the difference
$r^{(p)}-{\bar r}$ approaches zero as $w^2$: \be r^{(p)}(w)-{\bar
r}(w)=O(w^2)\;.\ee Therefore, the ratio ${  {(r^{(p)}} - {{\bar
r})}  }/{w^2}$ is finite as $\omega$ tends to zero, showing that
${\tilde F}^{(\rm wall)}_{\perp ij}$ is regular at $w=0$.

\subsection{The case of two plane-parallel slabs}

In this Section we calculate the Green's functions for the case of
a cavity constituted by two non-magnetic homogeneous, isotropic
and spatially non-dispersive plane-parallel slabs separated by
vacuum. We assume that the slabs can be characterized by the
respective electric permittivities, $\epsilon_1(w)$ and
$\epsilon_2(w)$. We choose our cartesian coordinate system in such
a way that slab one occupies the region $-\infty <z\le 0$, while
slab two occupies the region $d \le z < \infty$, $d$ being the
separation between the two slabs. The formulae derived in the
preceding Section, for the one slab case, can be easily
generalized to the two slabs setting, on the basis of the
intuitive physical picture of the free Green's functions as
consisting of left and right moving waves originating from ${\bf
r}'$.

Let us consider first the scalar Green's function ${\tilde { G}}$.
Analogously to what we did in the previous Section, inside the
cavity (i.e. for $0  \le z,z'  \le d$) we set: \be {\tilde {
G}}^{(\rm cav)}({\bf r},{\bf r}',w)={\tilde { G}}^{(0)}({\bf
r}-{\bf r}')+\,{\tilde { F}}^{(\rm cav)}({\bf r}_{\perp}-{\bf
r}_{\perp}',z,z',w)\;.\label{decscacav}\ee The expression that one
finds for ${\tilde { F}}^{(\rm cav)}$ is analogous to ${\tilde {
F}}^{(\rm wall)}$, but  of course one must take account now of the
possibility of multiple reflections off the two slabs. This is
easily done, by  inserting for each reflection by slab $i$ the
appropriate reflection coefficients ${\bar r}_i(w)$, that has an
expression analogous to Eq. (\ref{refsca}) (with $\epsilon_i(w)$
in the place of $\epsilon (w)$). Moreover, a factor $e^{-2
k_{\perp} d}$ must be included for each round-way trip from one
slab to the other and back.  One obtains:
$$ {\tilde {F}}^{({\rm cav})} = \int \frac{d^2 {\bf k}_{\perp}}{2
\pi \,k_{\perp}}\left[ \left(\frac{1}{{\cal A}}-1
\right)\left(e^{i {\bar {\bf k}}^{(+)}\cdot ({\bf r}-{\bf r}')} +
e^{i {\bar {\bf k}}^{(-)}\cdot ({\bf r}-{\bf r}')}\right)
\right.$$ \be\left.-\frac{1}{{\cal A}}\left( {\bar r}_1 \,e^{i
{\bar {\bf k}}^{(+)}\cdot {\bf r} -i {\bar {\bf k}}^{(-)}\cdot{\bf
r}'}+{\bar r}_2\,e^{i  {\bar {\bf k}}^{(-)}\cdot {\bf r} -i {\bar
{\bf k}}^{(+)}\cdot{\bf r}'-2 k_{\perp} d}
\right)\right]\;,\label{Fcav}\ee where ${\cal A}=1-{\bar
r}_1(w)\,{\bar r}_2(w)\,\exp (-2 k_{\perp}\,d)$.

For the transverse Green's function, we set: \be {\tilde {\bf
G}}^{(\rm cav)}_{\perp}({\bf r},{\bf r}',w)={\tilde {\bf
G}}^{(0)}_{\perp}({\bf r}-{\bf r}',w)+{\tilde {\bf F}}^{(\rm
cav)}_{\perp}({\bf r}_{\perp}-{\bf
r}_{\perp}',z,z',w)\;,\label{dectencav}\ee with \be {\tilde {\bf
F}_{\perp}}^{({\rm cav})}={\tilde {\bf U}}^{({\rm cav})}+{\tilde
{\bf V}}^{({\rm cav})}\;,\label{Gcavsplit}\ee where the symbols
have the obvious meaning, analogously to previous Section. The
same arguments that led us to write Eq. (\ref{Fcav}) now give:
\begin{widetext}$$ {\tilde { U}_{ij}}^{({\rm cav})}= i \int \frac{d^2 {\bf k}_{\perp}}{{2
\pi}{k_z}} \left\{ \left[ \left(\frac{1}{{\cal A}_s}-1
\right)\left(e^{i { {\bf k}}^{(+)}\cdot ({\bf r}-{\bf r}')} + e^{i
{{\bf k}}^{(-)}\cdot ({\bf r}-{\bf r}')}\right) +
\frac{{r}_1^{(s)}}{{\cal A}_s}   \,e^{i { {\bf k}}^{(+)}\cdot {\bf
r} -i {{\bf k}}^{(-)}\cdot{\bf r}'}+ \frac{{r}_2^{(s)}}{{\cal
A}_s}\,e^{i { {\bf k}}^{(-)}\cdot {\bf r} -i { {\bf
k}}^{(+)}\cdot{\bf r}'+2 i k_z d}\right]   e_{\perp i} e_{\perp j}
\right.$$ $$ \left.+\frac{1}{k_0^2}\left[\left(\frac{1}{{\cal
A}_p}-1 \right)\left(\xi^{(+)}_i \xi^{(+)}_j e^{i { {\bf
k}}^{(+)}\cdot ({\bf r}-{\bf r}')} +\xi^{(-)}_i \xi^{(-)}_j e^{i {
{\bf k}}^{(-)}\cdot ({\bf r}-{\bf r}')}\right) + \xi^{(+)}_i
\xi^{(-)}_j\,\frac{r_{1}^{(p)}}{{\cal A}_p}  \,e^{i {\bf
k}^{(+)}\cdot {\bf r}-i{\bf k}^{(-)}{\bf r}' } \right.
\right.$$\be \left.\left. + \,\xi^{(-)}_i
\xi^{(+)}_j\,\frac{r_{2}^{(p)}}{{\cal A}_p} \,e^{i {\bf
k}^{(-)}\cdot {\bf r}-i{\bf k}^{(+)}{\bf r}' + 2 i k_z
d}\frac{}{}\right]\right\}\;,\label{Ucav} \ee
\end{widetext}
where $r_i^{(\alpha)},\;\alpha=s,p$ are the Fresnel reflection
coeffcients of slab $i$ for polarization $\alpha$, and ${\cal
A}_{\alpha}=1-r_1^{(\alpha)}r_2^{(\alpha)}\,\exp(2 i k_z d)$. For
${\tilde {\bf V}}^{({\rm cav})}$ we obtain: \be {\tilde
{V}}_{ij}^{({\rm cav})}=\frac{1}{k_0^2} \frac{\partial^2 {
\Psi}^{({\rm cav})}}{\partial x_i \partial x'_j}\;,\label{Vcav}\ee
where
$$ {\tilde {\Psi}}^{({\rm cav})} = \int \frac{d^2 {\bf
k}_{\perp}}{2 \pi \,k_{\perp}}\left[ \left(\frac{1}{{\cal A}}-1
\right)\left(e^{i {\bar {\bf k}}^{(+)}\cdot ({\bf r}-{\bf r}')} +
e^{i {\bar {\bf k}}^{(-)}\cdot ({\bf r}-{\bf r}')}\right)\right.$$
\be \left. - \frac{1}{{\cal A}}\left( {\bar r}_1 \,e^{i {\bar {\bf
k}}^{(+)}\cdot {\bf r} -i {\bar {\bf k}}^{(-)}\cdot{\bf r}'}+{\bar
r}_2\,e^{i  {\bar {\bf k}}^{(-)}\cdot {\bf r} -i {\bar {\bf
k}}^{(+)}\cdot{\bf r}'-2 k_{\perp} d}
\right)\right]\;.\label{psicav}\ee Again we find, as in one slab
case, that the expression of ${\tilde {\Psi}}^{({\rm cav})}$
coincides with that of ${\tilde {F}}^{({\rm cav})}$. The same
considerations used in the one-slab case can be now repeated for
${\tilde {F}}^{({\rm cav})}$ and ${\tilde {\bf F}}^{({\rm
cav})}_{\perp}$ to show that both quantities are analytic in
${\cal C}^+$, and have a finite limit for vanishing $w$.

\section{COMMUTATION RELATIONS FOR THE EM FIELDS INSIDE A CAVITY}

In this Section we compute the quantities $A^{(\rm mat)}$,
$A_{\perp ij}^{({\rm mat})}$, $B_{\perp ij}^{({\rm mat})}$ and
$C_{\perp ij}^{({\rm mat})}$ for the two slab setting considered
in the previous Section. The corresponding quantities shall be
denoted by $A^{(\rm cav)}$, $A_{\perp ij}^{({\rm cav})}$,
$B_{\perp ij}^{({\rm cav})}$ and $C_{\perp ij}^{({\rm cav})}$,
respectively. We shall see that they all vanish, as a consequence
of the analyticity and fall-off properties at large frequencies of
the reflection coefficients of all real materials. As seen in Sec.
III, vanishing of these quantities entails that the e.m. field
satisfies free-field commutation relations in the empty region
between the slabs.

Consider first the quantity $A^{(\rm cav)}({\bf r},{\bf r}')$.
From its definition Eq. (\ref{Amat}) it follows that  $A^{(\rm
cav)}$ can be expressed in terms of  ${\tilde {F}}^{({\rm cav})}$
as: \be A^{(\rm cav)}({\bf r},{\bf r}')=\lim_{\tau \rightarrow
0^+} \int_{\Gamma} \frac{d w}{2 \pi}{\tilde {F}}^{({\rm
cav})}({\bf r},{\bf r}',w)\, e^{-i w\,\tau}\;,\label{F0}\ee
where ${\tilde {F}}^{({\rm cav})}$ is given in Eq. (\ref{Fcav}).
In Appendix B it is shown that the reflection coefficient ${\bar
r}(w)$ of any real material vanishes like $w^{-2}$ for large
values of $|w|$ and this implies, as can be seen by inspection of
Eq. (\ref{Fcav}), that ${\tilde {F}}^{({\rm cav})}$ approaches
zero like $w^{-3}$. Therefore ${\tilde {F}}^{({\rm cav})}$ is
absolutely integrable, and then in Eq. (\ref{F0}) we can take the
$\tau$-limit inside the integral. After we do it we obtain: \be
A^{(\rm cav)}({\bf r},{\bf r}')= \int_{\Gamma} \frac{d w}{2
\pi}{\tilde {F}}^{({\rm cav})}({\bf r},{\bf r}',w)\,
.\label{F0bis}\ee  The $w^{-3}$ fall-off rate of  $F^{(\rm cav)}$
at infinity now permits to close the integration contour in Eq.
(\ref{F0bis}) in the upper complex $w$-plane ${\cal C}^+$, and
then analyticity of ${\tilde {F}}^{({\rm cav})}$  in ${\cal C}^+$
implies at once that the integral is zero. Therefore we conclude
\be A^{(\rm cav)}({\bf r},{\bf r}')=0\;.\ee

We turn now to the quantity $A^{(\rm cav)}_{\perp ij}({\bf r},{\bf
r}')$. In view of its definition Eq. (\ref{Aijmat}) we have \be
A^{(\rm cav)}_{\perp ij}({\bf r},{\bf r}')=\lim_{\tau \rightarrow
0^+} \int_{\Gamma} \frac{d w}{2 \pi}{\tilde {F}}^{({\rm
cav})}_{\perp ij}({\bf r},{\bf r}',w)\, e^{-i w\,\tau}\;,\ee and
then to prove that it vanishes, we need consider the fall-off
properties of ${\tilde F}_{\perp ij}^{({\rm cav})}$. According to
Eq. (\ref{Gcavsplit}), it is the sum of two terms: ${\tilde
F}_{\perp ij}^{({\rm cav})}={\tilde U}_{\perp ij}^{({\rm
cav})}+{\tilde V}_{\perp ij}^{({\rm cav})}$. As to ${\tilde
V}_{\perp ij}^{({\rm cav})}$ we see, by inspection of Eqs.
(\ref{Vcav}) and (\ref{psicav}), that the fall-off rate of ${\bar
r}(w)$ implies that ${\tilde V}_{\perp ij}^{({\rm cav})}$
falls-off like $w^{-4}$. Consider now ${\tilde U}_{\perp
ij}^{({\rm cav})}$. We note first that, because of the $k_z$
factors in the exponentials, all terms in the r.h.s. of Eq.
(\ref{Ucav}) decay exponentially fast as $w$ goes to infinity in
${\cal C}^+$ along any direction not parallel to the real axis.
Along the real axis, since Fresnel reflection coefficients of all
real materials decay  like $w^{-2}$ (see Appendix B), ${\tilde
U}_{\perp ij}^{({\rm cav})}$ decays at least as fast as $w^{-3}$
(in fact a more careful analysis carried out in Appendix C shows
that the rate of decay is actually $w^{-4}$). Therefore, ${\tilde
F}_{\perp ij}^{({\rm cav})}$ decays in all directions in ${\cal
C}^+$ at least like $w^{-3}$ and then, by following exactly the
same reasoning  used in the case of $ A^{(\rm cav)}({\bf r},{\bf
r}')$ we can prove that \be A^{(\rm cav)}_{\perp ij}({\bf r},{\bf
r}')=0\label{Fcaveti}\;.\ee We remark that the above equation
holds also when either ${\bf r}$ or ${\bf r}'$ or both belong to
the slabs surfaces.  Consider now the quantity $B_{\perp
ij}^{({\rm cav})}({\bf r},{\bf r}')$. Recalling its definition Eq.
(\ref{Bijmat}), we have \be B^{(\rm cav)}_{\perp ij}({\bf r},{\bf
r}')=i\,\lim_{\tau \rightarrow 0^+} \int_{\Gamma} \frac{d w}{2
\pi}\,w\,{\tilde {F}}^{({\rm cav})}_{\perp ij}({\bf r},{\bf
r}',w)\, e^{-i w\,\tau}\;.\label{Bijbis}\ee Thanks to the $w^{-3}$
fall-off rate of ${\tilde F}_{\perp ij}^{({\rm cav})}(w)$, the
extra power of $w$   does not spoil convergence of the
$w$-integral on the r.h.s. of Eq. (\ref{Bijbis}), and therefore
the same arguments used to prove that $A^{(\rm cav)}_{\perp ij}$
is zero can be used to obtain \be B_{\perp ij}^{({\rm cav})} ({\bf
r},{\bf r}' ) =0\:. \label{onetderFij}\ee Finally, we consider the
quantity $C_{\perp ij}^{(\rm cav)}({\bf r},{\bf r}')$. For this we
have \be C_{\perp ij}^{(\rm cav)}({\bf r},{\bf r}')= \,\lim_{\tau
\rightarrow 0^+} \int_{\Gamma} \frac{d w}{2 \pi}\,w^2\,{\tilde
{F}}^{({\rm cav})}_{\perp ij}({\bf r},{\bf r}',w)\, e^{-i
w\,\tau}\;.\label{Cij}\;.\ee Proving that $C_{\perp ij}^{(\rm
cav)}$ vanishes requires much more labor, because of the {\it two}
extra powers of $w$   in the integrand on the r.h.s. of Eq.
(\ref{Cij}). We relegate the proof  in Appendix C, where we show
that the decay rate of ${\tilde F}_{\perp ij}^{({\rm cav})}$ is
actually $w^{-4}$, which is sufficiently fast to imply: \be
C_{\perp ij}^{(\rm cav)}({\bf r},{\bf r}')=0\;.\ee Having proved
that the quantities $A^{(\rm cav)}$, $A_{\perp ij}^{(\rm cav)}$,
$B_{\perp ij}^{(\rm cav)}$ and $C_{\perp ij}^{(\rm cav)}$ vanish,
we then reach the important conclusion that in the empty space
between two dielectric and/or conducting slabs the e.m. fields
satisfy free-field equal-time commutation relations, Eq.
(\ref{canUU}-\ref{charEE}). This result is consistent with what
was expected on the basis of the microscopic theory, for a system
of non-relativistic charged particles interacting with the e.m.
field, as we have seen in Sec. II. We remark that no singularities
are encountered as ${\bf r}$ and ${\bf r}'$ approach the slabs
surfaces, and therefore the canonical form of the free-space
commutators also holds on the surfaces of the slabs. It is
important to realize that these results are intimately tied to
analyticity and fall-off properties of the reflection coefficients
of real materials.

\section{COMMUTATION RELATIONS OUTSIDE IDEAL CONDUCTORS}

In this Section we investigate the commutation relations satisfied
by the e.m. fields outside ideal conductors. Ideal conductors are
characterized by the fact that they have ${\it constant}$
reflection coefficients. Indeed, by taking the limit $\epsilon
\rightarrow \infty$ in Eqs. (\ref{refsca}), (\ref{rs}) and
(\ref{rp}), we find that for an ideal conductor ${\bar r}$ and
$r^{(p)}$ are one, and $r^{(s)}$ is minus one at all frequencies.
Obviously, constant reflection coefficients are analytic in ${\cal
C}^+$, and therefore the main difference between ideal conductors
and real ones is that the reflection coefficients of the former
{\it do not} vanish in the limit of large frequencies. We shall
see below that this feature  entails that the e.m. field outside
the conductors, and on their surfaces, fail to satisfy free-field
canonical equal-time commutation relations.

In order to determine the commutation relations satisfied by the
e.m. field we consider again Eqs.
(\ref{etcomcavUU}-\ref{etcomcavEE}) that remain valid also for
ideal conductors. All that we have to do then is to evaluate the
quantities on the r.h.s. of these equations, using the values of
the reflection coefficients pertaining to ideal conductors. We
consider first the simpler case of a single conducting slab.

We start by evaluating the quantity ${ F}^{(\rm id\,wall)}$, where
the superscript $({\rm id \; wall})$ stands for a slab made of an
ideal metal. From Eq. (\ref{Fwall0}) we note that for ${\bar
r}(w)=1$, ${\tilde F}^{(\rm id\, wall)}$ becomes independent of
the frequency and, upon taking the inverse time-Fourier transform,
one easily finds that $F^{(\rm id\,wall)}$ is proportional to
$\delta(t-t')$. Then $F^{(\rm id\,wall)}$ is zero for all $t>t'$
and therefore from Eq. (\ref{gencomUU}) we have \be\langle[U({\bf
r},t),U({\bf r}',t')]\rangle =0\;\;\;\;\;{\rm
(ideal\;conductors)}. \ee  Upon taking account also of Eq.
(\ref{gencomUA})   we see that outside an ideal conductor, all
two-times commutators involving the scalar potential $U$  have
vanishing expectation values, and this implies
$$
U({\bf r},t)\equiv 0\;\;\;\;\;{\rm (ideal\;conductors)}.
$$
Therefore, outside an ideal conducting slab the longitudinal
electric field is zero. We evaluate now the quantity ${F}^{(\rm
id\; wall)}_{\perp ij}$. Upon using the identity \be -e_{\perp i}
e_{\perp j}+\frac{\xi_i^{(+)} \xi_j^{(-)}}{k_0^2}=-\lambda_k
\delta_{ik}\delta_{jk}+\frac{k_i^{(+)} k_j^{(-)}}{k_0^2}\;,\ee
where $\lambda_1=\lambda_2=-\lambda_3=1$, one finds that, for
${\bar r}=r^{(p)}=1$ and $r^{(s)}=-1$, Eqs.(\ref{Gwallsplit}),
(\ref{Ufin}) and (\ref{Psiwall0}) lead to:
$$
{\tilde F}^{(\rm id\; wall)}_{\perp ij}=-\int \frac{d^2 {\bf
k}_{\perp}}{2 \pi}\left[\frac{i}{k_z}\left(\lambda_k
\delta_{ik}\delta_{jk}- \frac{k_i^{(+)} k_j^{(-)}}{k_0^2}  \right)
\right.$$\be \left.\times
e^{ik_z(z+z')}+\frac{1}{k_{\perp}}\frac{{\bar k}_i^{(+)} {\bar
k}_j^{(-)}}{k_0^2}e^{-k_{\perp}(z+z')} \right]e^{i {\bf k}_{\perp}
\cdot ({\bf r}_{\perp}-{\bf r}'_{\perp})}\;. \ee  By a similar
computation as the one described in Appendix A, it is  possible to
verify that the r.h.s. of the above equation can also be written
in the following form: $$ {\tilde F}^{(\rm id\;wall)}_{\perp
ij}=-\int \frac{d^3 {\bf k}}{2 \pi^2} \frac{1}{k^2-k_0^2}
\left(\lambda_k \delta_{ik}\delta_{jk}- \frac{k_i k_j^{(r)}}{k^2}
\right)$$ \be \times \,e^{i[{\bf k}_{\perp}\cdot({\bf
r}_{\perp}-{\bf r}'_{\perp})+k_3(z+z')]}  \;, \label{Fijidtra}\ee
where ${\bf k}^{(r)}={\bf k}_{\perp}-k_3 {\hat {\bf z}}$.   From
this expression we see that in the case of an ideal wall ${\tilde
F}^{(\rm id\,wall)}_{\perp ij}$ decays for large frequencies only
like $w^{-2}$, and not like $w^{-4}$ as we found in the case of a
slab made of a real material. This fall-off rate is sufficient to
prove, by the same steps used in the previous Section, that  Eq.
(\ref{Fcaveti}) remains valid. Therefore we find that also in the
case of an ideal slab the equal-time commutators for the vector
potential have the canonical form Eq. (\ref{freeAA}) at all points
outside the slab, including its surface. The $w^{-2}$ fall-off
rate is not sufficient however to ensure validity of   Eq.
(\ref{onetderFij}), and we show now that for an ideal conductor
  Eq. (\ref{onetderFij}) indeed fails to be
true. To see this  we take the inverse Fourier transform of Eq.
(\ref{Fijidtra}), as defined in Eq. (\ref{invtra}). The frequency
integral, for $t>t'$ can be easily evaluated by closing the
contour $\Gamma$ in the lower complex plane (which is possible now
because the   r.h.s. of Eq. (\ref{Fijidtra}) is analytic also
there), and by noticing that the integrand  has poles only at
$k_0=\pm k$. We get:
$$ {F}^{(\rm id\;wall)}_{\perp ij}=-c\,\int \frac{d^3 {\bf k}}{2
\pi^2\,k} \left(\lambda_k \delta_{ik}\delta_{jk}- \frac{k_i
k_j^{(r)}}{k^2} \right)$$ \be \times \,e^{i[{\bf
k}_{\perp}\cdot({\bf r}_{\perp}-{\bf
r}'_{\perp})+k_3(z+z')]}\,\sin[c k (t-t')]\;.\label{Fwallct} \ee
Then, from   Eq. (\ref{etcomcavAE}) we obtain:
$$ \langle[A_{\perp i}({\bf r},t),E_{\perp j}({\bf
r}',t)]\rangle$$ \be=-4 \pi i \,\hbar \,c\,
[\delta_{ij}^{\perp}({\bf r}-{\bf r}')-\;
 \delta^{({\rm id \;wall})}_{ij}({\bf r}_{\perp}-{\bf r}'_{\perp},z+z')]\,,\label{etcomcavAEid}\ee
where we defined $$  \delta^{({\rm id\;wall})}_{ij}({\bf
r}_{\perp}-{\bf r}'_{\perp},z+z')=\int \frac{d^3 {\bf k}}{(2
\pi)^3} \left(\lambda_k \delta_{ik}\delta_{jk}- \frac{k_i
k_j^{(r)}}{k^2} \right)$$ \be \times \,e^{i[{\bf
k}_{\perp}\cdot({\bf r}_{\perp}-{\bf
r}'_{\perp})+k_3(z+z')]}\;,\label{deltawall}\ee We note that
$\delta^{({\rm id\;wall})}_{ij}$ is a smooth function for $z+z'>0$
approaching zero for large $z$ and $z'$, but it is singular when
both $z$ and $z'$ belong to the slab surface (i.e. for $z=z'=0$).
In particular, for $i=j=1$, Eq. (\ref{etcomcavAEid}) gives
$$ \langle[A_{\perp
1}({\bf r},t),E_{\perp 1}({\bf r}',t)]\rangle$$ \be=-4 \pi i
\,\hbar \,c\, [\delta_{11}^{\perp}({\bf r}-{\bf r}')-\;
 \delta_{11}^{\perp}({\bf r}_{\perp}-{\bf r}'_{\perp},z+z')]\,,\label{etcomcavAEidxx}\ee
in agreement with the finding of Ref. \cite{milonni}. By using Eq.
(\ref{comcavEEperp}), and Eq. (\ref{Fwallct}), it is easy to
verify that the canonical commutation relations for the components
of the transverse electric field, Eq. (\ref{freeEE}) remain valid.

We turn now to the more elaborate case of two plane-parallel ideal
slabs. We shall be brief here, the analysis being similar to the
one slab case. First we note that, similarly to ${\tilde F}^{(\rm
id\,wall)}$, also the quantity ${\tilde F}^{(\rm id\,cav)}$
becomes independent of the frequency when perfectly reflecting
slabs are considered, as it is easily seen  from Eq. (\ref{Fcav}).
Therefore, $F^{(\rm id\, cav)}$ is proportional to $\delta(t-t')$,
and again we conclude the the scalar potential can be taken to be
zero outside the slabs. We consider now the transverse Green's
function. A somewhat lengthy, but straightforward  computation
analogous to the one done for the one-slab case, gives the
following expression for the quantity ${F}^{(\rm id\;cav)}_{\perp
ij}$
$$ {F}^{(\rm
id\;cav)}_{\perp ij}=-c\,\int \frac{d^3 {\bf k}}{2 \pi^2\,k}
\frac{1}{{\cal A}^{({\rm id})}}\left[\left( \delta_{ij}- \frac{k_i
k_j}{k^2} \right)e^{i k_3(2 d+z-z')}\right.$$
$$+ \left( \delta_{ij}- \frac{k_i^{(r)}
k_j^{(r)}}{k^2} \right)e^{i k_3(2 d-z+z')}+\left(\lambda_k
\delta_{ik}\delta_{jk}- \frac{k_i k_j^{(r)}}{k^2} \right)$$
$$\times e^{i
k_3(z+z')}\left.+\left(\lambda_k \delta_{ik}\delta_{jk}-
\frac{k_i^{(r)} k_j}{k^2} \right)e^{i k_3(2d-z-z')}\right]$$ \be
\times \,e^{i{\bf k}_{\perp}\cdot({\bf r}_{\perp}-{\bf
r}'_{\perp}) }\,\sin[c k (t-t')]\;,\label{Fcavct} \ee where ${\cal
A}^{({\rm id})}=1-\exp(2i k_3 d)$. By using this Equation, and
recalling Eqs. (\ref{etcomcavAA}) and (\ref{etcomcavEE}), we
easily see that the equal-time commutators for the vector
potential on one hand and for the transverse electric field on the
other, both vanish inside the cavity and on the slabs surfaces, in
agreement with the free-field case,  Eq. (\ref{freeAA}) and Eq.
(\ref{freeEE}). On the other hand, from Eq. (\ref{etcomcavAE}) we
get:
$$ \langle[A_{\perp i}({\bf r},t),E_{\perp j}({\bf
r}',t)]\rangle$$ \be=-4 \pi i \,\hbar \,c\,
[\delta_{ij}^{\perp}({\bf r}-{\bf r}')-\;
 \delta^{({\rm id \;cav})}_{ij}({\bf r}_{\perp}-{\bf r}'_{\perp},z,z')]\,,\label{etcomcavbisAEid}\ee
 where $$  \delta^{({\rm id\;cav})}_{ij}({\bf
r}_{\perp}-{\bf r}'_{\perp},z,z')=\int \frac{d^3 {\bf k}}{(2
\pi)^3}\frac{1}{{\cal A}^{({\rm id})}}\left[\left( \delta_{ij}-
\frac{k_i k_j}{k^2} \right)\right.$$
$$\times e^{i k_3(2 d+z-z')}+ \left( \delta_{ij}- \frac{k_i^{(r)}
k_j^{(r)}}{k^2} \right)e^{i k_3(2 d-z+z')}$$
$$+\left(\lambda_k
\delta_{ik}\delta_{jk}- \frac{k_i k_j^{(r)}}{k^2} \right) e^{i
k_3(z+z')}$$ \be\left. +\left(\lambda_k \delta_{ik}\delta_{jk}-
\frac{k_i^{(r)} k_j}{k^2} \right)e^{i k_3(2d-z-z')}\right]
\,e^{i{\bf k}_{\perp}\cdot({\bf r}_{\perp}-{\bf r}'_{\perp}) }
\;.\label{deltacav} \ee We note that the first and the second
terms between the square brackets on the r.h.s. of  Eq.
(\ref{deltacav}) represent smooth functions of $z$ and $z'$ at all
points between the slabs, including their surfaces, while the
third and fourth terms  are singular, respectively, on the surface
of slab one (i.e. for $z=z'=0$) and slab two (i.e. for $z=z'=d$).
Moreover, we observe that in the limit of large separations $d$,
and for fixed $z$ and $z'$, the phase factors involving $d$ in the
first, second and fourth terms between the square brackets on the
r.h.s. of  Eq. (\ref{deltacav}),  oscillate more and more rapidly,
and so suppress the corresponding terms. In this limit
$\delta^{({\rm id \;cav})}_{ij}$ tends to $\delta^{({\rm id
\;wall})}_{ij}$,  and then Eq. (\ref{etcomcavbisAEid}) reproduces
Eq. (\ref{etcomcavAEid}).

From the above analysis, we see that while all other commutators
have the free-field form, the presence of the extra terms
$\delta^{({\rm id\;wall})}_{ij}$ and $\delta^{({\rm
id\;cav})}_{ij}$ on the r.h.s. of Eqs. (\ref{etcomcavAEid}) and
(\ref{etcomcavbisAEid}) respectively, implies that perfect-mirror
b.c.  lead to equal-time commutation relations  for the vector
potential and the electric field of a different form from the
free-field ones, Eq. (\ref{freeAE}). In the one-slab case, Eqs.
(\ref{etcomcavAEid}) and (\ref{deltawall}) show that the
free-field form of the commutators is recovered only when the
quantity $ \delta^{({\rm id\;cav})}_{ij}$ can be neglected, and
this occurs   at points $z$ and $z'$ that are far from the slab.
In the cavity setting, Eqs. (\ref{etcomcavbisAEid}) and
(\ref{deltacav}) show that free-field commutation relations are
recovered only provided that the quantities $2 d +z-z'$,
$2d-z+z'$, $z+z'$ and $2d-z-z'$ are simultaneously large. This is
only possible for large cavities, and for points $z$ and $z'$ far
form both conductors.   This is in contrast with what was found in
the previous Section, where we proved that in the case of real
materials free-fields equal-time canonical commutation relations
retain their validity everywhere between the slabs, including on
their surfaces.

\section{CONCLUDING REMARKS}

In this paper we have determined the commutation relations
satisfied by the quantized e.m. field outside one or two
plane-parallel dielectric and/or conducting slabs in vacuum,
assuming that the slabs are made of   isotropic and homogeneous,
spatially non-dispersive materials, with arbitrary
frequency-dependent dispersion and absorption. Using a general
form of macroscopic quantum electrodynamics, we have found that at
all points between the slabs, including on their surfaces, the
e.m. field satisfies canonical commutation relations of the same
form  as in empty space, in full agreement with the microscopic
theory. This result is a general consequence of analyticity and
fall-off properties at large frequencies satisfied by the
reflection coefficients of all real materials.

We have also shown that free-field equal-time commutation
relations do not obtain outside one or two conducting slabs, if
the latter are modelled as perfect mirrors, because  of extra
terms that appear in the commutator of the vector potential with
the electric field. Free-field commutators are only recovered  at
points that are sufficiently far from the mirror. In the one-slab
setting, our findings coincide with those obtained by Milonni
\cite{milonni} in his investigation on the Casimir effect. Since
no such deviation from  free-field commutators is found in the
case of real materials, we draw the conclusion that the modified
form  of the field commutation relations  implied by
perfect-mirror b.c. is an artifact of this idealized model, as it
was conjectured to be the case in Ref.\cite{barnett} on the basis
of a simplified one-dimensional form of QED. Even if the
commutator of the vector potential with the electric field, being
a gauge dependent quantity, is not a physically observable
quantity, failure of perfect mirror b.c. to reproduce the correct
free-field form of the equal-time commutation relations near the
surfaces of the conductors, indicates that a certain amount of
caution should be used when these idealized b.c. are used in
investigations of proximity phenomena originating from the
quantized e.m. field in the presence of conductors.

Before closing, we would like to comment on   possible
generalization of the results derived  in this paper to other
materials, including magnetic materials, and  non isotropic or
spatially dispersive media. Consideration of  isotropic magnetic
materials offers no difficulties, because it just requires
substituting  in our formulae the well-known expression for
Fresnel reflection coefficients,  for a medium with magnetic
permeability $\mu $. On the other hand, it is   known today that
the general formulae Eq. (\ref{gencomUU}-\ref{gencomAA})
expressing the expectation values of the field commutators   in
terms of their classical Green's functions, are valid for
arbitrary media \cite{bimonte,pitaevskii}, and therefore one can
use them also in the case of anisotropic and/or spatially
dispersive media provided only that one is able to determine the
reflection coefficients for a slab made of these materials. Since
reflection coefficients of all media are analytic functions of the
complex frequency in the upper complex plane, and fall off to zero
at large frequencies \cite{Wooten}, it is therefore  expected that
(free-space) canonical commutation relations remain valid also for
these more general materials.

\appendix

\section{WEYL REPRESENTATION OF THE TRANSVERSE GREEN'S FUNCTION IN EMPTY SPACE}

As it is well known, the time Fourier-transform of the Green's
function for the transverse e.m. field in free space is given by
the formula: \be {\tilde {G}}^{(0)}_{\perp ij}=\int \frac{d^3 {\bf
k}}{2 \pi^2} \frac{1}{k^2-k_0^2}\left(
\delta_{ij}-\frac{k_i\,k_j}{k^2}\right)  \;e^{i{\bf k}\cdot ({\bf
r}-{\bf r}')}\;. \label{gthree}\ee   An expression analogous to
Weyl's  representation of the scalar Green's function, Eq.
(\ref{weyl}),  can be obtained by performing the integral over
$k_3$ in Eq.(\ref{gthree}). The integral can be done easily by
suitably closing the $k_3$ contour of integration in the complex
$k_3$ plane (for $z \ge z'$ one closes the contour in upper
half-plane, for $z \le z'$ in the lower plane). The integral then
receives contributions from   poles in the integrand of
Eq.(\ref{gthree}), which arise from two sources. The first one is
the factor involving the inverse of $k^2-k_0^2$, which gives rise
to poles at $k_3=\pm k_z$, where $k_z=\sqrt{k_0^2-k_{\perp}^2}$
(the square root is defined such that ${\rm Im}(k_z) > 0$).
Importantly, the second source of poles is the gauge fixing term
inside the round brackets, proportional to the inverse of $k^2$,
which has poles at $k_3=\pm i k_{\perp}$.  We correspondingly
split ${\tilde {\bf G}_{\perp}}^{(0)}$ as the sum of two terms:
\be {\tilde {\bf G}_{\perp}}^{(0)}={\tilde {\bf U}}^{(0)}+{\tilde
{\bf V}}^{(0)}\;,\ee where ${\tilde {\bf U}}^{(0)}$ accounts for
the former set of poles, and ${\tilde {\bf V}}^{(0)}$ for the
latter. We find:
 \be {\tilde { U}_{ ij}}^{(0)}=\frac{i}{2 \pi} \int \frac{d^2
{\bf k}_{\perp}}{{k_z}} \left(\delta_{ij}-\frac{k_i^{(\pm)}
k_j^{(\pm)}}{k_0^2} \right)    \,e^{i {\bf k}^{(\pm)}\cdot ({\bf
r}-{\bf r}') }\;,\label{U}\ee where ${\bf k}^{(\pm)}={\bf
k}_{\perp}\pm k_z \hat{{\bf z}}$, and \be {\tilde {V}_{ ij}}^{(0)}
= \frac{1}{{k_0^2 }}\int \frac{d^2 {\bf k}_{\perp}}{{2
\pi}\,k_{\perp}} {\bar{k}_i^{(\pm)}\bar{k}_j^{(\pm)}}\,
 \,e^{i {\bar {\bf k}}^{(\pm)}\cdot ({\bf r}-{\bf
r}')}\;,\label{V}\ee where  ${\bar {\bf k}}^{(\pm)}={\bf
k}_{\perp}\pm i\,k_{\perp} \hat{{\bf z}}$, and in both Eqs.
(\ref{U}) and  (\ref{V}) the upper (lower) sign is for $z \ge z'$
($z \le z'$). It is now convenient to further transform the
expression of ${\tilde {U}_{ ij}}^{(0)}$ by considering the
following decomposition of the identity $\delta_{ij}$: \be
\delta_{ij}=e_{\perp i} e_{\perp j}+\frac{1}{k_0^2}(\xi^{(\pm)}_i
\xi^{(\pm)}_j+ k_i^{(\pm)}k_j^{(\pm)})\;,\label{deco}\ee where
${\bf e_{\perp}}={\hat {\bf z}}\times {\hat {\bf k}}_{\perp}$, and
${\bf \xi}^{\pm}=k_{\perp}{\hat {\bf z}} \mp k_z {\hat {\bf
k}}_{\perp}$. Upon replacing $\delta_{ij}$, in Eq. (\ref{U}), by
the r.h.s. of Eq. (\ref{deco}), we obtain:
 \be {\tilde { U}_{ ij}}^{(0)}=\frac{i}{2 \pi} \int \frac{d^2
{\bf k}_{\perp}}{{k_z}} \left( e_{\perp i} e_{\perp
j}+\frac{\xi^{(\pm)}_i \xi^{(\pm)}_j}{k_0^2}  \right) \,e^{i {\bf
k}^{(\pm)}\cdot ({\bf r}-{\bf r}') }\;.\label{Ubis}\ee As for
${\tilde {V}_{ ij}}^{(0)}$, we note that it represents a pure
scalar contribution, for it can be written as: \be {\tilde {V}_{
ij}}^{(0)}=\frac{1}{k_0^2} \frac{\partial^2 {\tilde
\Psi}}{\partial x_i \partial x'_j}\;,\label{Vbis}\ee where \be
{\tilde \Psi}=\int \frac{d^2 {\bf k}_{\perp}}{{2 \pi}\,k_{\perp}}
 \,
 \,e^{i {\bar {\bf k}}^{(\pm)}\cdot ({\bf r}-{\bf
r}')}\;.\label{Psi}\ee Upon combining Eq. (\ref{Ubis}) and Eq.
(\ref{Vbis}), we obtain the following final expression for
${\tilde {G}}^{(0)}_{\perp ij}$:
$${\tilde {G}}^{(0)}_{\perp ij}=\frac{i}{2 \pi} \int \frac{d^2 {\bf
k}_{\perp}}{{k_z}} \left( e_{\perp i} e_{\perp
j}+\frac{\xi^{(\pm)}_i \xi^{(\pm)}_j}{k_0^2}  \right)  \,e^{i {\bf
k}^{(\pm)}\cdot ({\bf r}-{\bf r}') }$$ \be+\,\frac{1}{k_0^2}\,
\frac{\partial^2 {\tilde \Psi}}{\partial x_i
\partial x'_j}\;.\label{Gzerotra}\ee

\section{PROPERTIES OF THE REFLECTION COEFFICIENTS}

In this Appendix, we briefly review some   general properties of
the reflection coefficients of   dielectrics and conductors, that
are important for the present paper. They are a direct consequence
of the general properties of the electric permittivity
$\epsilon(\omega)$. As it is well known \cite{lifs2}, the
permittivity $\epsilon(\omega)$ of a causal medium is an analytic
function of the complex frequency $w$ in the upper complex plane
${\cal C}^+$. Moreover, its imaginary part $\epsilon''$ is never
zero in ${\cal C}^+$, except along the positive imaginary axis
($w= i \xi\;,\xi>0$), where $\epsilon$ is positive and
monotonically decreasing. For large (complex) frequencies $w$, the
electric permittivities of all materials approach one, and have an
asymptotic expansion of the form \cite{lifs2}: \be
\epsilon(w)=1-\frac{A}{w^2}+i\frac{B}{w^3}+\cdots\;,\label{asref}\ee
where $A$ and $B$ are real positive constants characteristic of
the material.

As a consequence of the above analyticity properties of
$\epsilon(w)$, the reflection coefficients ${\bar r}$, $r^{(p)}$
and $r^{(s)}$, given by Eqs.(\ref{refsca}), (\ref{rp}) and
(\ref{rs}) respectively, are   analytic functions in ${\cal C}^+$,
and they are   real and positive along the positive imaginary axis
\cite{Wooten}. The   asymptotic behavior of $\epsilon(w)$ implies
that for large frequencies  ${\bar r}(w)$ and $r^{(\alpha)}(w,{
k}_{\perp}),\;\alpha=s,p$   have asymptotic expansions of the
form: \be {\bar r}(w)\simeq \frac{{\bar C}}{w^2}+i\frac{{\bar
D}}{\omega^3}+\cdots\;,\; r^{(\alpha)}(w,{ k}_{\perp}) \simeq
\frac{C^{(\alpha)}}{w^2}+i\frac{D^{(\alpha)}}{w^3}+\cdots\;,\label{asrefbis}\ee
where ${\bar C}$, ${\bar D}$, $C^{(\alpha)}$ and ${\bar
D}^{(\alpha)}$ are real numbers characteristic of the material. We
remark  that $C^{(\alpha)}$ and ${\bar D}^{(\alpha)}$ are
independent of ${ k}_{\perp}$.

It is useful to consider also the behavior of the reflection
coefficient in the limit of zero frequency. In the case of a
dielectric, $\epsilon(w)$ approaches a positive constant
$\epsilon_0>1$ at zero frequency, and therefore for the reflection
coefficients ${\bar r}(0)$, $r^{(p)}(0)$ and $r^{(s)}(0)$ we find:
\be {\bar
r}(0)=r^{(p)}(0)=\frac{\epsilon_0-1}{\epsilon_0+1}<1\;,\;\;\;({\rm
dielectrics})\ee \be r^{(s)}(0)=0\;\;\;({\rm dielectrics}).\ee
Consider now  conductors. At sufficiently low frequency, the
permittivity of a conductor is of the form \be \epsilon(w)=4 \pi
i\,\frac{ \sigma_0}{w}\;,\;\;\;({\rm conductors})\ee where
$\sigma_0$ is ohmic conductivity. Then for the reflection
coefficients at zero frequency we find: \be {\bar
r}(0)=r^{(p)}(0)= 1\;,\;\;\;({\rm conductors})\ee \be
r^{(s)}(0)=0\;\;\;({\rm conductors}).\ee It is also useful to
estimate the behavior of the difference between $r^{(p)}(w)$ and
${\bar r}(w))$, as $w$ approaches zero. In the case of
dielectrics, one finds: \be r^{(p)}-{\bar
r}=\frac{\epsilon_0(\epsilon_0-1)}{(1+\epsilon_0)^2}\frac{w^2}{c^2
k_{\perp}^2}+O(w^4)\;, \ee while, in the case of conductors, we
find: \be r^{(p)}-{\bar r}= \frac{w^2}{c^2 k_{\perp}^2}+\frac{i
w^3}{c^2 k_{\perp}^2 \sigma_0}\left(\frac{3}{4 \pi}+\frac{\pi
\sigma_0^2}{c^2 k_{\perp}^2} \right)+O(w^4)\;.\ee We see that in
both cases, the difference $r^{(p)}-{\bar r}$ approaches zero as
$w^2$.

\section{PROOF THAT $C_{\perp ij}^{(\rm cav)}=0$}

In this Appendix we prove that the quantity $C_{\perp ij}^{(\rm
cav)}$ defined in Eq. (\ref{Cij}) is zero.

Upon recalling Eq. (\ref{dectencav}), we first split $C_{\perp
ij}^{(\rm cav)}$  as \be C_{\perp ij}^{(\rm cav)}({\bf r},{\bf
r}')=D_{ij}({\bf r},{\bf r}')+E_{ij}({\bf r},{\bf r}')\;,\ee where
\be D_{ij}({\bf r},{\bf r}')=\lim_{t\rightarrow t'^+}\frac{
\partial^2 U_{ij}^{({\rm cav})}}{\partial t \,\partial t'}({\bf r},{\bf
r}',t-t') \label{Dij}\ee and \be E_{ij}({\bf r},{\bf
r}')=\lim_{t\rightarrow t'^+}\frac{
\partial^2 V_{ ij}^{({\rm cav})}}{\partial t \,\partial t'}({\bf r},{\bf
r}',t-t') \;.\label{Ezero} \ee In terms of Fourier transforms, we
can write the above quantities as: \be D_{ij}=\lim_{\tau
\rightarrow 0^+} \int_{\Gamma} \frac{d w}{2 \pi}\, w^2\,{\tilde
{U}}^{({\rm cav})}_{ij}({\bf r},{\bf r}',w)\,e^{-i w
\tau}\;.\label{Dij0}\ee \be E_{ij}=\lim_{\tau \rightarrow 0^+}
\int_{\Gamma} \frac{d w}{2 \pi}\, w^2\,{\tilde {V}}^{({\rm
cav})}_{ij}({\bf r},{\bf r}',w)\,e^{-i w \tau}\;.\label{Eij0}\ee
By the same arguments used earlier in the case $ A^{(\rm
cav)}({\bf r},{\bf r}')$, we see that $D_{ij}$ and $E_{ij}$
vanish, provided that ${\tilde {U}}^{({\rm cav})}_{ij}$ and
${\tilde {V}}^{({\rm cav})}_{ij}$ fall off like $w^{-4}$ or
faster. By inspection of Eqs. (\ref{Vcav}) and (\ref{psicav}), and
recalling that ${\bar r}(w)$ falls off like $w^{-2}$, we can
easily see that this is the case for ${\tilde V}_{\perp ij}^{({\rm
cav})}$. Therefore we have: \be E_{ij}({\bf r},{\bf r}')=0\;.\ee
Consider now   the quantity ${\tilde {U}}^{({\rm cav})}_{ij}$
whose expression is provided by Eq. (\ref{Ucav}).
In order to estimate the fall-off rate of the various terms in the
r.h.s. of Eq. (\ref{Ucav}),  we need recall that Fresnel
reflection coefficients of all real materials decay like $w^{-2}$
(see Appendix B). Since the quantities $(1/{\cal A}_{s}-1)$ and
$(1/{\cal A}_{p}-1)$ then decay like $w^{-4}$, we see that all
terms involving these quantities in the r.h.s. of Eq. (\ref{Ucav})
fall off at least like $w^{-5}$ at large frequencies, and
therefore they can be neglected. Consider now the remaining terms
in the expression for ${\tilde {U}}^{({\rm cav})}_{ij}({\bf
r},{\bf r}',w)$.   Upon noticing that for large $w$,
$\xi^{(\pm)}_i \xi^{(\pm)}_j=-\xi^{(\pm)}_i
\xi^{(\mp)}_j=-\xi^{(\mp)}_i \xi^{(\pm)}_j=w^2\,{\hat k}_{\perp i}
{\hat k}_{\perp j}+O(w)$, we get: \begin{widetext}$$ {\tilde
{U}}^{({\rm cav})}_{ij}= i \int \frac{d^2 {\bf k}_{\perp}}{{2
\pi}{k_z}}\left[\left({{r}_1^{(s)}} \,e^{i { {\bf k}}^{(+)}\cdot
{\bf r} -i {{\bf k}}^{(-)}\cdot{\bf r}'}+ {{r}_2^{(s)}}\,e^{i {
{\bf k}}^{(-)}\cdot {\bf r} -i { {\bf k}}^{(+)}\cdot{\bf r}'+2 i
k_z d}\right)\right. e_{\perp i}e_{\perp j}$$ \be -\left.\left(
{{r}_1^{(p)}}    \,e^{i { {\bf k}}^{(+)}\cdot {\bf r} -i {{\bf
k}}^{(-)}\cdot{\bf r}'}+ {{r}_2^{(p)}}\,e^{i { {\bf k}}^{(-)}\cdot
{\bf r} -i { {\bf k}}^{(+)}\cdot{\bf r}'+2 i k_z d}\right)k_{\perp
i}k_{\perp j}\; \right]+O(w^{-4})\label{C7}\ee
\end{widetext}
Having reached this point, we  take advantage of the fact that to
order $w^{-3}$ included, both Fresnel coefficients are independent
of $k_{\perp}$ (see Eq. (\ref{asrefbis})). By virtue of this, the
r..h.s. of Eq. (\ref{C7}) can be written as: \be {\tilde
{U}}^{({\rm cav})}_{ij}=\frac{1}{w^2}\sum_{\alpha=s,p}\sum_{m=1,2}
{K^{(\alpha)}_{m|ij}} I_m (w)+O(w^{-4})\label{Ulead} \ee where
${K^{(\alpha)}_{m|ij}}$ are material dependent constants, and
  \be I_1(w)=i\int \frac{d^2 {\bf
k}_{\perp}}{{2 \pi}{k_z}}\,e^{i { {\bf k}}^{(+)}\cdot {\bf r} -i
{{\bf k}}^{(-)}\cdot{\bf r}'} \;,\ee while
 \be I_2(w)=i\int \frac{d^2 {\bf
k}_{\perp}}{{2 \pi}{k_z}}\, e^{i { {\bf k}}^{(-)}\cdot {\bf r} -i
{ {\bf k}}^{(+)}\cdot{\bf r}'+2 i k_z d}\;.\ee It is a simple
matter to check that the integrals $I_i$ can be also written as
\be I_1(w)=\int \frac{d^3{\bf k}}{2 \pi^2}
\,\frac{1}{k^2-k_0^2}\,e^{i [{ {\bf k}}_{\perp}\cdot ({\bf
r}_{\perp} -{\bf r}_{\perp}')+k_3(z+z')]}\;, \ee and \be
I_2(w)=\int \frac{d^3{\bf k}}{2 \pi^2} \,\frac{1}{k^2-k_0^2}\,e^{i
[{ {\bf k}}_{\perp}\cdot ({\bf r}_{\perp} -{\bf
r}_{\perp}')+k_3(2d-z-z')]}\;.\ee From this we see that both $I_1$
and $I_2$ fall-off like $w^{-2}$, and therefore, in view of Eq.
(\ref{Ulead}), we find that ${\tilde {U}}^{({\rm cav})}_{ij}$
decays like $w^{-4}$ or faster. Therefore $D_{ij}$ is zero, and
then we obtain the desired result: \be  C_{\perp ij}^{(\rm
cav)}({\bf r},{\bf r}')=0\;.\ee

\end{document}